\documentclass{emulateapj}

\usepackage{float}
\usepackage{amsmath}
\usepackage{epsfig,floatflt}
\usepackage{subfigure}

\def\na{New\ Astron.\ }

\def\para#1{\medskip\noindent{\it #1\/---}}
\def\microns{\ifmmode \,\mu$m$
             \else \,$\mu$m\fi}

\begin{document}

\title{CMB component separation by parameter estimation}

\author{H. K. Eriksen\altaffilmark{1}, C. Dickinson\altaffilmark{2},
  C. R. Lawrence\altaffilmark{3}, C. Baccigalupi\altaffilmark{4},
  A. J. Banday\altaffilmark{5}, K. M. G\'orski\altaffilmark{6},
  F. K. Hansen\altaffilmark{7}, P.\ B.\ Lilje\altaffilmark{8},
  E. Pierpaoli\altaffilmark{9}, M. D. Seiffert\altaffilmark{10},
  K. M. Smith\altaffilmark{11} and K. Vanderlinde\altaffilmark{12}}

\altaffiltext{1}{Institute of Theoretical Astrophysics, University of
Oslo, P.O.\ Box 1029 Blindern, N-0315 Oslo, Norway; Centre of
Mathematics for Applications, University of Oslo, P.O.\ Box 1053
Blindern, N-0316 Oslo; Jet Propulsion Laboratory, M/S 169/327, 4800
Oak Grove Drive, Pasadena CA 91109; California Institute of
Technology, Pasadena, CA 91125; email: h.k.k.eriksen@astro.uio.no}

\altaffiltext{2}{Department of Astronomy, California Institute of
Technology, 1200 E. California Blvd., MS105-24, Pasadena, CA, 91125;
email: cdickins@astro.caltech.edu}

\altaffiltext{3}{JPL, M/S 169/327, 4800 Oak Grove Drive, Pasadena CA
91109; email: charles.r.lawrence@jpl.nasa.gov}

\altaffiltext{4}{SISSA/ISAS, Astrophysics Sector, Via Beirut, 4,
I-34014 Trieste; INFN, Sezione di Trieste, Via Valerio 2, I-34014
Trieste, Italy; Institut f\"ur Theoretische Astrophysik, Universit\"at
Heidelberg, Albert-ˆúberle-Strasse 2, D-69120 Heidelberg, Germany;
email: bacci@sissa.it}

\altaffiltext{5}{Max-Planck-Institut f\"ur Astrophysik,
Karl-Schwarzschild-Str.\ 1, Postfach 1317, D-85741 Garching bei
M\"unchen, Germany; email: banday@MPA-Garching.MPG.DE}

\altaffiltext{6}{Jet Propulsion Laboratory, M/S 169/327, 4800 Oak
Grove Drive, Pasadena CA 91109; Warsaw University Observatory, Aleje
Ujazdowskie 4, 00-478 Warszawa, Poland; California Institute of
Technology, Pasadena, CA 91125; email:
Krzysztof.M.Gorski@jpl.nasa.gov}

\altaffiltext{7}{Institute of Theoretical Astrophysics, University of
Oslo, P.O.\ Box 1029 Blindern, N-0315 Oslo, Norway; email:
f.k.hansen@astro.uio.no} 

\altaffiltext{8}{Institute of Theoretical Astrophysics, University of
Oslo, P.O.\ Box 1029 Blindern, N-0315 Oslo, Norway; Centre of
Mathematics for Applications, University of Oslo, P.O.\ Box 1053
Blindern, N-0316 Oslo; email: per.lilje@astro.uio.no}

\altaffiltext{9}{Theoretical Astrophysics, MC 130-33, California
  Institute of Technology, Pasadena, CA 91125; email: pierpa@caltech.edu}

\altaffiltext{10}{Jet Propulsion Laboratory, 4800 Oak Grove Drive,
Pasadena CA 91109; email: Michael.D.Seiffert@jpl.nasa.gov}

\altaffiltext{11}{Department of Physics, University of Chicago,
  Chicago IL 60637; email: kmsmith@uchicago.edu}

\altaffiltext{12}{Kavli Institute for Cosmological Physics and Enrico
Fermi Institute, University of Chicago, Chicago, IL, 60637; email:
kvand@uchicago.edu}


\begin{abstract}
We propose a solution to the CMB component separation problem based on
standard parameter estimation techniques.  We assume a parametric
spectral model for each signal component, and fit the corresponding
parameters pixel by pixel in a two-stage process.  First we fit for
the full parameter set (e.g., component amplitudes and spectral
indices) in low-resolution and high signal-to-noise ratio maps using
MCMC, obtaining both best-fit values for each parameter, and the
associated uncertainty. The goodness-of-fit is evaluated by a $\chi^2$
statistic. Then we fix all non-linear parameters at their
low-resolution best-fit values, and solve analytically for
high-resolution component amplitude maps.  This likelihood approach
has many advantages: The fitted model may be chosen freely, and the
method is therefore completely general; all assumptions are
transparent; no restrictions on spatial variations of foreground
properties are imposed; the results may be rigorously monitored by
goodness-of-fit tests; and, most importantly, we obtain reliable error
estimates on all estimated quantities.  We apply the method to
simulated \emph{Planck} and six-year \emph{WMAP} data based on
realistic models, and show that separation at the $\mu\textrm{K}$
level is indeed possible in these cases.  We also outline how the
foreground uncertainties may be rigorously propagated through to the
CMB power spectrum and cosmological parameters using a Gibbs sampling
technique.
\end{abstract}


\maketitle

\section{Introduction}

As experimental techniques improve rapidly, and new high-sensitivity
ground-based, balloon-borne, and satellite missions are being planned and
implemented, the main problem in CMB measurement has changed from instrumental
noise to separation of the cosmological CMB signal from non-cosmological
foreground signals.  This problem will become even more important as our focus
shifts from observations of temperature anisotropies to polarization
anisotropies; while a simple template-fitting approach proved adequate for the
first-year \emph{WMAP} analysis
\citep{bennett:2003a,bennett:2003b,hinshaw:2003}, no such hopes can be held for
future polarization experiments.

While component separation is a difficult problem, it is not intractable.  Since
the cosmological CMB radiation follows a virtually perfect black body spectrum
\citep{mather:1999}, whereas all known non-cosmological signals have
non-thermal spectra, it should be possible to disentangle the various
contributions using spectral information.  This fact motivated multiple
frequencies on the \emph{COBE}-DMR experiment (three bands between 31 and
90\,GHz), the current \emph{WMAP} experiment (five bands between 23 and 94\,GHz),
and the future \emph{Planck} experiment (nine bands between 30 and 857\,GHz).

While the necessity of multi-frequency observations has been
recognized within the cosmological community for a long time, there
has been uncertainty about how those observations should be utilized.
Many different methods have been proposed, including the Maximum
Entropy Method
\citep{barreiro:2004,bennett:2003b,hobson:1998,stolyarov:2002,stolyarov:2005},
the Internal Linear Combination method
\citep{bennett:2003b,tegmark:2003,eriksen:2004a}, Wiener filtering
\citep{bouchet:1999,tegmark:1996}, and the Independent Component
Analysis method \citep{maino:2002,maino:2003,donzelli:2005}. Some of
these methods (e.g., Baccigalupi et al.\ 2004 and Stivoli et al. 2005)
have been applied to polarization data. These methods all have their
origins in the field of computational image processing.

In this paper, we advocate a more direct approach to the component
separation problem, following in the footsteps of \citet{brandt:1994}.
We choose a physically motivated parametric model for each significant
signal component, and fit the free parameters by means of
well-established algorithms, such as MCMC and non-linear
searches. (For an alternative spectral matching algorithm, see
Delabrouille, Cardoso \& Patanchon 2003.) The advantages of this
approach are many: it does not require priors (although it can benefit
from them); no assumptions about the spatial structure of the
foreground properties are imposed; the method scales proportionally to
the number of pixels; the results may be verified by means of
goodness-of-fit tests; and, most importantly, the method yields
accurate uncertainties for all estimated quantities.

We begin by developing a simple algorithm that is able to
analyze real-world data in the presence of realistic noise.  Whereas
\citet{brandt:1994} relied exclusively on non-linear fitting, and was
therefore quite unstable with respect to noise, we take advantage of
more recent developments (most importantly Markov Chain Monte Carlo)
to make the algorithm both more robust, and also to produce accurate
errors.  We also suggest a procedure for propagating these results into
final data products, namely the CMB power spectrum and cosmological
parameters, by means of a Gibbs sampling method. 

\begin{figure*}

\begin{center}
\mbox{\epsfig{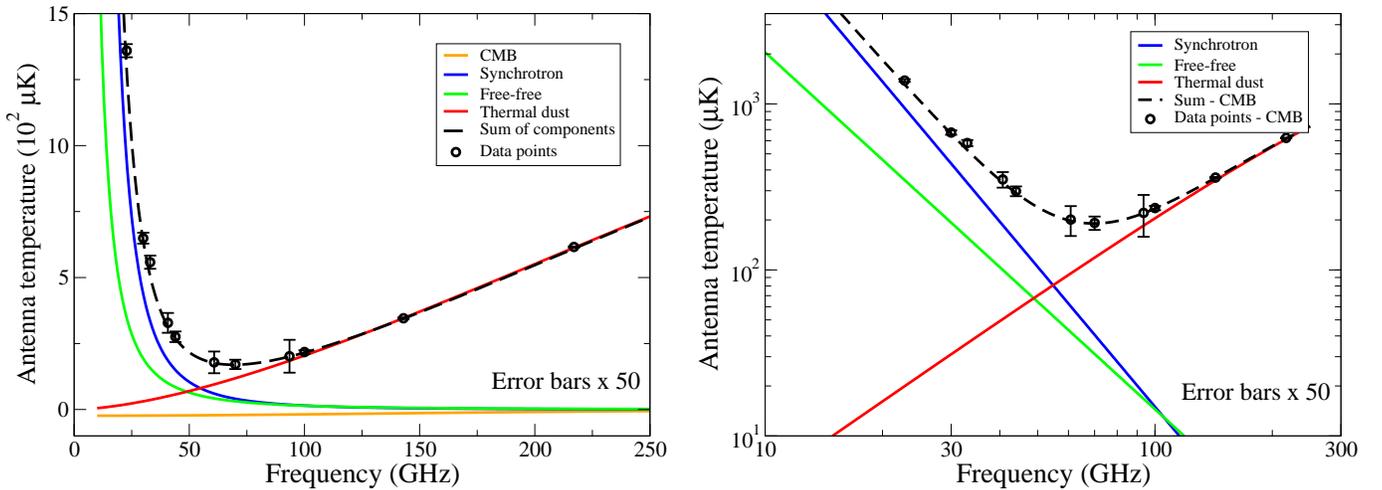}}
\end{center}

\caption{Component separation using multi-frequency measurements
  (linear units in left panel, logarithmic units in right panel). Most
  signal components has a well-defined frequency spectrum that may be
  parametrized by one or a few parameters, and component separation
  may therefore be viewed as a standard parameter estimation problem.
  The example shown here is based on one single pixel in a simulated
  data set corresponding to the six-year \emph{WMAP} and the
  \emph{Planck} experiments, as discussed in \S\,\ref{sec:planck}. The
  error bars on the data points are multiplied by a factor of 50 in
  order to make them visible on this scale. (Due to modeling errors,
  this particular fit has a $\chi^2$ of 44, and with five degrees of
  freedom, it is rejected at the 99.9999\% confidence level.)
}
\label{fig:fitting_illustration}
\end{figure*}

After establishing the algorithm, we apply it to realistic simulations
corresponding to the future \emph{Planck} and six-year \emph{WMAP} data, taking
into account the predicted noise distributions of each experiment.  We attempt
to model the foregrounds as accurately as possible, given our current
understanding of the involved foregrounds.

\section{Likelihood formulation of the component separation problem}

Assume that the observed data take the form of a multi-frequency set of sky
maps, $\mathbf{d}_{\nu}$, each of which may be written in the form
\begin{equation}
\mathbf{d}_{\nu} = \mathbf{A}\mathbf{s}_{\nu} + \mathbf{n}_{\nu},
\label{eq:model}
\end{equation}
where $\nu$ identifies frequency bands and sky maps 1 through $N$,
$\mathbf{s}_{\nu}$ is the true sky signal at the corresponding frequency band,
$\mathbf{n}_{\nu}$ is instrumental noise, and $\mathbf{A}$ denotes convolution
with the instrument beam.

We assume the noise component $\mathbf{n}_{\nu}$ to be Gaussian distributed
with vanishing mean and variance $\sigma^2_{\nu}(p)$, where $p$ is the pixel
number. Thus, the noise is uncorrelated both between pixels and between
frequency channels, but spatial variations in the variance are
allowed. 

Note that there is no frequency index on the beam operator in
Equation~\ref{eq:model}, indicating that all channels are assumed to have the
same beam response. For multi-resolution experiments, this implies that the sky
maps must be smoothed to a common resolution prior to analysis.  Equation
\ref{eq:model} is no longer strictly valid, since the noise is then correlated;
however, in practice it works reasonably well to approximate the noise term as
uncorrelated between pixels, with rms levels determined by Monte Carlo
simulations of processed noise.

The signal $\mathbf{s}_{\nu}$ may be decomposed into a sum of
components, $\mathbf{s}_{\nu} = \sum_{i} \mathbf{s}_{\nu}^{i}$, in
which the most important ones are the cosmological CMB signal and
three galactic foregrounds; synchrotron, free-free and dust
emission. Compact (unresolved) galactic and extra-galactic sources
could also be included in the list, but these are more conveniently
detected by other methods, such as wavelets (e.g., Vielva et
al.~2003).  In this paper we consider compact source removal as a part
of the pre-processing stage, and assume that resolved sources are
either masked, median filtered or fitted prior to analysis. Thus, we
include only diffuse foregrounds in the following, but note that more
work is needed on this issue.

Assume also that the frequency spectrum of each signal component may be
parametrized by a small number of free parameters, for instance an amplitude
and a spectral index $S = A \left(\nu/\nu_0\right)^{\beta}$.  Given a set of
multi-frequency CMB sky maps as described above, and a parametric signal model
$S_{\nu}(\theta)$ with free parameters $\theta$, we now establish both a point
estimate $\hat{\theta}$ for the free parameters, and the corresponding
uncertainties.  To do so, we use standard likelihood methods of parameter
estimation.

Since the noise is assumed to be Gaussian distributed and uncorrelated between
pixels, the likelihood $\mathcal{L}$ reduces to $\chi^2$, independent between
pixels\footnote{In practice this is not  strictly correct, since the CMB and
foreground components are indeed correlated between pixels, and we will at some
point work with smoothed sky maps, but it is a good approximation for component
separation purposes. Further, as described in \S\,\ref{sec:gibbs}, at least
spatial CMB correlations may be taken into account by Gibbs sampling.},
\begin{equation}
\ln \mathcal{L} = -\frac{1}{2} \sum_{\nu=1}^{N} \left(\frac{d_{\nu} -
  S_{\nu}(\theta)}{\sigma_{\nu}}\right)^2 = -\frac{1}{2} \chi^2.
\label{eq:chisq}
\end{equation}
The problem is thus reduced to mapping out a likelihood by some
numerical technique, for instance grid computation, MCMC, or
non-linear searches.

We illustrate the procedure in Figure~\ref{fig:fitting_illustration}, where the
results from an analysis of one arbitrarily chosen pixel are shown (see
\S\,\ref{sec:planck} for details).  The observed data points are marked by black
circles, and the fitted components are shown as smooth, colored curves. The
dashed black curve shows the sum of all components.  Four signal components are
included in this model, CMB, synchrotron, free-free, and thermal dust emission.
The parametric models for the foregrounds are perfect power laws for synchrotron
and free-free emission, and a one-component model for dust (see
Equation~\ref{eq:dust}).

Although we focus on temperature anisotropies in the present paper, the method
can handle polarization anisotropies equally well. In that case, the $\chi^2$
takes the form
\begin{equation}
\chi^2 = \sum_{\nu=1}^{N} 
\left(\bar{d}_{\nu} - \bar{S}_{\nu}(\theta)\right)^{\textrm{T}} 
\mathbf{N}_{\nu}^{-1}
\left(\bar{d}_{\nu} - \bar{S}_{\nu}(\theta)\right),
\end{equation}
where
\begin{equation}
\bar{d}_{\nu} = 
\left(
\begin{array}{c}
d^{I}_{\nu} \\
d^{Q}_{\nu} \\
d^{U}_{\nu}
\end{array}
\right),
\quad
\bar{S}_{\nu}(\theta) = 
\left(
\begin{array}{c}
S^{I}_{\nu}(\theta) \\
S^{Q}_{\nu}(\theta) \\
S^{U}_{\nu}(\theta)
\end{array}
\right)
\end{equation}
are the Stoke's $I$, $Q$, and $U$ parameters for the data and model
respectively, and $\mathbf{N}_{\nu}$ is the $3\times3$ $(I,Q,U)$ noise
correlation matrix for the pixel. (We still assume uncorrelated noise between
pixels and frequency bands, but not between the three Stoke's parameters for
each individual pixel.)

\section{Parametric Model Fits}
\label{sec:model}

In this section we introduce the parametric model that we fit to the data in the
likelihood analysis, starting with a review of the currently favored parametric
signal models $\mathbf{s}_{\nu}^i$ for each signal component.  We emphasize that
the procedure as such is general, and any parametric model may be included in
the analysis.  Then we discuss how to take into account the effect of non-zero
instrumental bandwidths, and discuss implementation details that lead to more
transparent computer code.  Note that the parametric signal models we adopt for
\emph{fitting} the various components may differ from the parametric models
that are used for \emph{modeling} the components later in section 
\ref{sec:fgs}, as will certainly be the case in dealing with real data.

\subsection{Signal components}

\para{CMB anisotropies}
The cosmological background component is, due to its black body nature
\citep{mather:1999}, most easily characterized by its thermodynamic temperature
$T_{\textrm{CMB}}$, or equivalently, the anisotropy temperature $\Delta
T_{\textrm{CMB}} = T_{\textrm{CMB}} - T_{0}$, where $T_0 = 2.725\,\textrm{K}$
\citep{fixsen:2002} is the average CMB temperature. However, as discussed
later, the foreground components are more easily described in terms of antenna
temperatures, and therefore we choose to convert the CMB signal accordingly.
The CMB signal model then reads
\begin{equation}
s_{\textrm{CMB}}(\nu) = \Delta T_{\textrm{CMB}} \frac{x^2 e^x}{(e^x -
1)^2},
\end{equation}
where $x = h\nu/kT_0$, $h$ is the Planck constant, and $k$ is
Boltzmann's constant. 

\para{Synchrotron emission}
Synchrotron emission from the Galaxy originates in relativistic cosmic ray (CR)
electrons spiraling in the Galactic magnetic field. The morphology of the
observed emission depends on the distribution of the relativistic electrons in
the Galaxy, and the Galactic magnetic field structure.  In the Galactic plane,
the latter exhibits a large-scale ordering with the field parallel to the
spiral arms (the regular component).  Superimposed on this is real small-scale
structure (the irregular component) which shows variations between the arm and
inter-arm regions and with gas phase.  The regular and irregular components seem
to be of comparable magnitude.  At high latitudes, there is a contribution from
the Galactic halo, and specific nearby structures (e.g., the North Polar Spur).
Variations in the frequency spectral index of the synchrotron continuum
emission arise from variations in the CR electron energy spectrum, which has a
range of distributions depending on age and the environment of origin (e.g.,
supernova explosions or diffuse shocks in the interstellar medium). 

The synchrotron emission may be accurately modeled by means of a simple power
law over a considerable range of frequencies,
\begin{equation}
s_{\textrm{s}}(\nu) = A_{\textrm{s}}\left(\frac{\nu}{\nu_{0,\textrm{s}}}\right)^{\beta_{\textrm{s}}}.
\label{synchrotron_scaling}
\end{equation}
Here $A_{\textrm{s}}$ is the synchrotron amplitude\footnote{Note that
italic font $A$ is used for component amplitudes, while bold font
$\mathbf{A}$ is used for beam convolution.}(measured in antenna
temperature $\mu\textrm{K}$) at some reference frequency
$\nu_{0,\textrm{s}}$, and $\beta_{\textrm{s}}$ is the synchrotron
spectral index.  Since the spectral index varies with both frequency
and position on the sky, at least two free parameters are required to
describe the synchrotron emission properly in a given direction.

\citet{lawson:1987} studied the spectral index variation based on low-frequency
radio surveys, and found that the brighter regions away from the Galactic plane
have typical values of $\beta$ at 100 and 800~MHz of 2.55 and 2.8,
respectively.  \citet{reich:1988} used radio continuum surveys of the Northern
sky at 408 and 1420~MHz to demonstrate a range of spectral index values
between 2.3 and 3.0, with a typical dispersion $\Delta \beta= \pm 0.15$.  The
steepest spectra were observed towards the North Polar Spur, and there was a
flattening in spectral index towards higher latitudes in the Galactic
anti-centre direction.  Such behavior has been confirmed over the full sky
by \citet{reich:2003}, who find that spectral flattening is particularly
pronounced in the Southern hemisphere

At higher frequencies, the brightness temperature spectral index is expected to
steepen by $\sim 0.5$ due to electron energy losses \citep{platania:1998}.
\citet{banday:2003} derived a mean spectral index between 408~MHz and
19.2\,GHz from the \citet{cottingham:1987} survey and between 31.5, 53, and
90\,GHz from the $COBE$-DMR data.  The steep spectral index of $\sim 3.1$
for Galactic latitudes $|b| > 15^\circ$ is consistent with expectations.
\citet{bennett:2003b} claim that the spectral break occurs near the K-band. 
Spectral indices above 10\,GHz are likely between 2.7 to 3.2.

\para{Free-free emission}Free-free emission is the brems\-strahlung radiation
resulting from the Coulomb interaction between free electrons and ions in the
Galaxy.  Free-free emission is weaker than synchrotron emission at frequencies
below $\sim 1$\,GHz over most regions of the sky, with exceptions in the
Galactic ridge and the brighter features of the local Gould Belt system
\citep{dickinson:2003}.  Free-free almost never dominates at high latitudes in
any frequency band, and is therefore difficult both to observe and to 
simulate.

%
%
Using the relations summarized in \citet{dickinson:2003}, it can be shown that
the brightness temperature of the free-free emission is described by the
relation
\begin{equation}
  T_{\textrm{ff,b}} \propto \nu^{-2}\, T_{\textrm{e}}^{-0.5}\, \left(
  \ln [0.04995\nu^{-1}] +  1.5 \ln T_{\textrm{e}} \right) 
\label{eq:freefree_scaling}
\end{equation}
where $T_{\textrm{e}}$ is the electron temperature.

\citet{shaver:1983} used radio recombination lines to show that the electron
temperature of H{\sc ii} regions at the galactocentric radius of the Sun
($R_{0}=8.5$ kpc) is $7200\pm1200$\,\hbox{K}.  \citet{paladini:2005} found
similar results from a larger sample containing many weaker sources.  At high
galactic latitudes, the ionized hydrogen typically will be within $\sim 1$\,kpc
of the Sun \citep{dickinson:2003}, and we expect the electron temperature to be
in the range $T_{\textrm{e}}=7000$--8000\,K, although it is possible that the
diffuse emission at a given Galactocentric distance may differ from that of the
higher density H{\sc ii} regions on the plane.

The frequency dependence is therefore well constrained, with an effective
spectral index of $\beta_{\textrm{ff}}=-2.14$ at the frequencies of interest,
the range over 10--100\,GHz being of order $-2.1$ to $-2.2$ and steepening still
further to $-2.3$ at hundreds of\,GHz. 

\citet{fink:2004a} has analyzed the \emph{WMAP} data and found a significant
component with a free-free like spectrum  within $30^\circ$ of the Galactic
center. The component is uncorrelated  with H$\alpha$ emission and may be
indicative of hot ($\sim 10^6\,\textrm{K}$) gas. We do not attempt to include
such a component here.

In this paper, we model free-free emission as a simple power law with a fixed
spectral index,
\begin{equation}
s_{\textrm{ff}}(\nu) = A_{\textrm{ff}}\left(\frac{\nu}{\nu_{0,\textrm{ff}}}\right)^{-2.14}.
\end{equation}
Thus only one parameter is required for free-free emission.  Future experiments
may need to estimate the electron temperature directly from the data.  In that
case Equation \ref{eq:freefree_scaling} should be used directly, at the cost of
introducing one extra free parameter into the fit. 

\para{Thermal dust emission}The thermal dust emission that contributes to the
frequencies of interest for CMB analysis arises from grains large enough to be in
thermal equilibrium with the interstellar radiation field, and is known from
analysis of the \emph{IRAS} and \emph{COBE}-DIRBE data to peak at a wavelength of
approximately 140\microns.  At higher frequencies, there is a contribution from
the optically active modes of PAH molecules, but these are not of interest here.

Currently preferred dust emission models \citep{fink:1999} extrapolate from
high-frequency \emph{COBE}-FIRAS and -DIRBE observations to CMB frequencies using
combinations of modified blackbody fits and accounting for dust temperature
variations.  Such fits approximate the integrated contributions to the emission
from multiple components of dust, i.e., with different grain properties (chemical
composition and size) and equilibrium temperatures.  The best-fit model (model~8
of Finkbeiner et al.~1999) assumes two main components:
\begin{equation}
s_{\textrm{d}}(\nu )=F\left(\frac{\nu}{\nu_{0,\textrm{d}}}\right)^{\beta_{\textrm{d}}(\nu )}.
\label{dust_scaling}
\end{equation}
Here $F$ represents the combined \emph{COBE}-DIRBE and \emph{IRAS} template
\citep{sfd98}, $\nu_{0,\textrm{d}}=3000$\,GHz, and $\beta_{\textrm{d}}(\nu)$ is
dependent on the frequency as discussed above,
\begin{equation}
\beta_{\textrm{d}}(\nu)=\frac{\log{\frac{d(\nu)}{d(\nu_{0,\textrm{d}})}}}{\log \frac{\nu}{\nu_{0,\textrm{d}}}}\ ,
\label{dust_spectral_index}
\end{equation}
\begin{equation}
\begin{split}
\,\,d(\nu)=\frac{q_{1}}{q_{2}}\,f_{1}
\left(\frac{\nu}{\nu_{0,\textrm{d}}}\right)^{3+\alpha_{1}}
\frac{1}{e^{\frac{h\nu}{kT_{1}}}-1}\,+\\
\quad\quad f_{2}\left(\frac{\nu}{\nu_{0,\textrm{d}}}\right)^{3+\alpha_{2}}
\frac{1}{e^{\frac{h\nu}{kT_{2}}}-1}.
\end{split}
\label{dust_scaling_function}
\end{equation}
with best-fit parameters $f_{1}=0.0363$, $q_{1}/q_{2}=13$,
$\alpha_{1}=1.67$, $\alpha_{2}=2.70$, $T_{1}=9.4$\,K, $T_{2}=16.2$\,K,
and $f_2 = 1-f_1$ \cite{fink:1999}.

In principle, these equations may serve as our parametric model for fitting the
dust emission spectrum.  However, few (current or future) CMB experiments have
sufficient power to constrain six parameters for dust alone, and simplifications
are therefore unavoidable.  Rather than fitting the full form as given above, we
therefore choose the simpler ``model 3'' of \citet{fink:1999}, setting $f_1 = 1$
and $T_1=18.1\,\textrm{K}$, but letting $\alpha_1$ vary freely. Equation
\ref{dust_scaling} may then be simplified to
\begin{equation}
s_{\textrm{d}}(\nu) = A_{\textrm{d}}
\frac{\nu}{e^{\frac{h\nu}{kT_1}}-1} \frac{e^{\frac{h\nu_{\textrm{d},0}}{kT_1}}-1}{\nu_{\textrm{d},0}}
\left(\frac{\nu}{\nu_{\textrm{d},0}}\right)^{\beta_{\textrm{d}}},
\label{eq:dust}
\end{equation}
where $A_{\textrm{d}}$ is the thermal dust amplitude at a reference
frequency $\nu_{\textrm{d},0}$, and a $\beta_{\textrm{d}}$ is a free parameter.
Thus, the fitted model is a power law modulated by a slowly decreasing
function of order unity over the frequencies of interest.

Such a parameterization does not allow for the spectral break that does exist in
the \emph{COBE}-FIRAS at approximately 500\,GHz, and which may reflect either the
emissivity of different grain components, frequency dependence of the emissivity
of the dominant grain component, or possibly a population of cold dust grains
mixed with the warmer dust \citep{reach:1995}.  However, at 500\,GHz the
signal of dust anisotropies is so strong compared with CMB anisotropies that it
is useless for purposes of component separation.

\para{Anomalous dust emission}Cross-correlation of the \emph{COBE}-DMR data with
the DIRBE map of thermal dust emission at 140\microns\ in \citet{kogut:1996}
revealed an anomalous component with rising spectrum from 53 to 31.5\,GHz.
\citet{banday:2003}, again using the DMR data together with a survey at 19.2\,GHz,
and independently the \emph{WMAP} team \citep{bennett:2003b}, suggested that this
component was well-described by a power-law spectrum with index $-2.5$ for
frequencies in the range $\sim$20--60\,GHz.  The latter proposed that the
emission originates in star-forming regions close to the Galactic plane.  However,
the favored model to explain this anomalous dust-correlated component is in
terms of the rotational emission from very small grains. 

\citet{draine:1998} have developed a three component model of this
'spinning dust' which contains contributions from the three phases of
the interstellar medium---the Cold Neutral Medium, the Warm Neutral
Medium, and the Warm Ionised Medium. The characteristic spectral
behavior of the model includes a rising spectrum up to a turn-over in
the range 10--20\,GHz, then a rapidly falling spectrum which can be
characterized by an effective spectral index in excess of 3 beyond
30\,GHz.  

Recent observations by \citet{fink:2004b} of dust correlated emission
outside of H{\sc ii} regions between 8 and 14\,GHz shows a rising
spectral slope and amplitude far exceeding that associated with
thermal dust emission.  More importantly, \citet{watson:2005} show
that observations of the Perseus molecular cloud made between 11 and
17\,GHz and augmented with the \emph{WMAP} data can be adequately
fitted by a spinning dust model.  Nevertheless, although the case is
compelling for describing the anomalous emission by such models, the
detailed morphology of the emission remains uncertain and no
unambiguous template to trace it exists.  It is also possible that the
emission may be confined to specific clouds at relatively low Galactic
latitude, leading to a more patchy distribution than for the diffuse
thermal dust contribution with some additional sensitivity to
environment.  Indeed, \citet{lagache:2003} presents evidence that the
excess emission is associated with small transiently heated dust
particles, which may be destroyed under certain physical conditions.
Given these uncertainties, we do not include this foreground component
in our studies.

\subsection{Non-zero bandwidths}

The previous sections describe the basic behavior of each signal component as a
function of well-defined frequencies.  However, real experiments integrate over
a range of frequencies (a ``frequency band''), typically with unequal weights, and
the observed signal strength does not equal that given by the central frequency
alone.

We take into account this effect through the concept of an ``effective
frequency'' $\nu_{\textrm{eff}}$, defined by 
\begin{equation}
S(\nu_{\textrm{eff}}) = \int f(\nu) S_{\nu} d\nu,
\label{eq:nu_eff}
\end{equation}
where $S_{\nu}$ is the frequency spectrum of the signal, and $f(\nu)$ is the
frequency response profile of the detector. Thus, the spectrum at the effective
frequency equals the average over the frequency band. The advantage of doing this
is simply that computationally expensive integrals are replaced by single point
computations.

In this paper we assume for simplicity that all frequency response functions
correspond to flat bandpass filters with sharp frequency cutoffs at $\nu_a$ and
$\nu_b$.  For simple power law models, such as those of synchrotron and free-free
emission as described above, the effective frequency of a signal component with
spectral index $\beta$ is then given by
\begin{equation}
\nu_{\textrm{eff}} = \left(\frac{1}{\beta+1} \frac{\nu_{b}^{\beta+1} -
  \nu_{a}^{\beta+1}}{\nu_b - \nu_a}\right)^{1/\beta}
\end{equation}
For more complicated spectra, Equation~\ref{eq:nu_eff} must be solved
numerically.  Fortunately, it is straightforward to pre-compute a grid of the
effective frequencies prior to the full analysis, since they only depend on the
frequency scalings and not the component amplitudes, and computational speed is 
not compromised.

\subsection{Implementation details}
\label{sec:implementation}

To simplify the computer code, it is convenient to introduce some general
notation.  For instance, if we can write all signal models in a common form, we do
not have to consider a list of special cases, but rather handle all cases with
the same code.

Indeed, all frequency spectra discussed above may be written in a common form,
namely that of a power law modulated by an arbitrary frequency-dependent function,
\begin{equation}
s_{\nu}(p) = \sum_{i=1}^{N_{\textrm{comp}}} A_{i}(p) c_{i,\nu_{\textrm{eff},i}}
\left(\frac{\nu_{\textrm{eff},i}}{\nu_{0,i}}\right)^{\beta_i(p)}.
\label{eq:unified_form}
\end{equation}
Here $A_i(p)$ and $\beta_i(p)$ are the ``free'' amplitude and index parameters
for component $i$ in each pixel $p$, and $c_{i,\nu_{\textrm{eff}_i}}$ is an
arbitrary function only dependent on frequency.

Specifically, synchrotron and free-free emission are included simply
by setting $c_{i,\nu} = 1$, while the CMB component is defined by
$\beta_{\textrm{CMB}} \equiv 0$ and $c_{\textrm{cmb},\nu} = x^2 e^x /
(e^x - 1)^2$, as discussed above. For dust, $c_{\textrm{d},\nu}$ is
given by equation \ref{eq:dust}.

Even anomalous dust could be included within this notation.  One
option is simply to tailor the correction factors $c_{i,\nu}$ to match
the predicted spinning dust spectrum \citep{draine:1998}, fix the
corresponding spectral index at zero, and then fit for the amplitude
only.  Another is to merge the spectrum with that of the thermal dust
emission, and thereby enforce identical spatial templates.

\section{Summary of previous results}

This paper may be seen as a natural continuation of the work started by
\citet{brandt:1994}, who considered how well future experiments could reconstruct
the CMB signal in presence of noise and foregrounds. Their approach, parameter
estimation, was the same as ours; however, they relied solely on maximum
likelihood estimation (i.e., non-linear fitting), and their results were
therefore less stable with respect to noise than the ones we present here, as
will be seen below.  Nevertheless, several of the conclusions drawn by
\citet{brandt:1994} are still valid for our work, and well worth repeating:
\begin{enumerate}

\item The number of frequencies must equal or exceed the number of
  fitted parameters, else the problem is mathematically degenerate.  This is
  obvious, but not trivial: no experiment to date has had the 
  minimum number of frequencies required to separate CMB, synchrotron, free-free,
  and dust fluctuations, even in their simplest form.

\item One should attempt to reduce the number of free parameters in the problem,
  as this gives greater stability with respect to noise. Seemingly gross
  simplifications, such as approximating both synchrotron and free-free emission by
  a single power law, can often yield improvements in the reconstruction.

\item It is usually advantageous to fit spectral parameters to
  reduced-resolution and low-noise data, and then solve for the
  amplitudes in the full-resolution data, fixing the indices at the
  smoothed values.

\item Due to the similarity between the synchrotron and free-free
  emission, better results are obtained whenever the latter is not a
  significant contaminant. Thus, if the free-free contamination could
  be constrained by radiation physics knowledge, it is well worth
  trying.

\end{enumerate}

While working on the present analysis, we have reproduced all of these
conclusions, and most of them have been taken into account when establishing the
prescription described below.  However, we will not elaborate further on these
issues here, but rather refer the interested reader to \citet{brandt:1994}, and
come back the above points as they are needed in the analysis.

\section{Method}
\label{sec:algorithm}

In this section we propose an algorithm for solving the parameter estimation
problem with sufficient speed and accuracy to be useful for practical analysis of
current and future data.  Each step of the algorithm consists of well-established
methods, and the approach should seem quite familiar.

\subsection{Overview}

The goal is to establish both a point estimate of all interesting parameters, and
their uncertainties.  Our prescription for doing so is as follows:
\begin{enumerate}
\item If required by noise levels or computational resources,
  downgrade all sky maps both in pixel and beam resolution.

\item For each low-resolution pixel,

  \begin{enumerate}

  \item choose a parametric model,

  \item solve for all parameters jointly by MCMC,

  \item estimate non-linear parameters and corresponding uncertainties
    by marginalizing over all other parameters,

  \item find the goodness-of-fit in terms of the $\chi^2$.

  \end{enumerate}

\item For each high-resolution pixel within a low-resolution pixel,

  \begin{enumerate}

  \item \emph{either} (fast but approximate analytical approach)

    \begin{enumerate}

    \item fix the non-linear parameters at the low-resolution pixel values,

    \item find maximum-likelihood estimates for all linear parameters
      (i.e., component amplitudes) by solving a linear equation,

    \item find corresponding uncertainties by analytic error
      propagation formula,

    \item estimate the goodness-of-fit in terms of the $\chi^2$ again.

    \end{enumerate}

  \item \emph{or} (exact but expensive Gibbs sampling approach)

    \begin{enumerate}

    \item use the low-resolution MCMC distributions to sample
    non-linear parameters numerically,

    \item given a set of non-linear parameters, sample amplitudes from
    their corresponding Gaussian distribution,

    \item given foregrounds, sample CMB sky map and power spectrum.

    \end{enumerate}

  \end{enumerate}

\end{enumerate}

The route outlined in step 3b) holds promise of a complete solution to the
foreground problem in a CMB context, since, if successful, the foreground
uncertainties are propagated all the way from noisy observations to the CMB power
spectrum and cosmological parameters.  However, in this paper we only present the
basic ideas, and leave the details for a future more comprehensive study.  All
high-resolution results presented in the following are thus based on the
analytical approach.

\subsection{Non-linear parameters and large-scale smoothing}

One of the main themes of \citet{brandt:1994} was the instability of a non-linear
fit with respect to noise.  This is not hard to understand.  In estimating
multiple parameters from a limited number of frequency bands, the maximum
likelihood point may easily slide along some degeneracy ridge on the likelihood
surface in the presence of realistic noise.  For all currently planned
experiments, additional degree-scale smoothing is a requirement in order to reach
reconstruction errors at the $\mu\textrm{K}$ level.

Another and more practical issue is the fact that our main algorithm relies on
MCMC analysis of each individual pixel.  This takes on the order of 100~seconds
per pixel.  Even though the algorithm parallelizes trivially because the pixels
are analyzed individually, a complete analysis at full
\emph{Planck} resolution ($\sim 50$ million pixels) would be unfeasible.

It is important to realize, however, that a full MCMC analysis is required only
for estimating non-linear parameters, such as spectral indices or dust
temperatures.  If all parameters in the problem are linear (i.e., component
amplitudes), an analytical computation is equally good.  We therefore compute the
complete probability distributions from reduced-resolution maps, fix the
high-resolution non-linear parameters at the corresponding low-resolution values,
and then solve for the high-resolution component amplitudes with alternative
methods, for instance analytically or by Gibbs sampling. 

When adequate data on foregrounds are in hand, the validity of this smoothing can
be tested.  If preliminary indications turn out to be true, and spectral indices
vary more slowly on the sky than amplitudes, the smoothing process will not lead
to significant loss of information.  In any case, the smoothing scale can be
optimized within the bounds of computational resources.

\subsection{Model selection}

A second main theme of \citet{brandt:1994} was the importance of model
selection.  They clearly demonstrated that a large number of parameters does not
necessarily yield a better CMB reconstruction.  Quite the contrary, too many
parameters often yield unphysical results.  In general, one should never fit more
parameters than required by the data.

In principle, it would be useful to have an automated prescription to identify
the optimal model for a given pixel.  To some extent, such a procedure is provided
by means of the so-called \emph{information criterion} (IC), an idea that was
introduced to CMB analysis by \citet{liddle:2004}.  The fundamental idea in this
approach is not to maximize the likelihood alone, but rather minimize the IC,
defined as follows,
\begin{equation}
\textrm{IC} = -2 \ln \mathcal{L} + \alpha k.
\label{eq:ic}
\end{equation}
Here $k$ is the number of parameters in the fit, and $\alpha$ is a penalty factor.
(Two useful choices for $\alpha$ are $\alpha_{\textrm{A}}=2$, the Akaike
information criterion, and $\alpha_{\textrm{B}} = \ln N$, the Bayesian information
criterion, $N$ being the number of data points.)  Within this framework, a new
parameter must prove its usefulness by returning a significant improvement in the
$\chi^2$ fit to be included in the model.

We implemented this approach in our codes, and obtained reasonable
results.  However, the model sky map had a clear tendency to be
patchy, and not necessarily well correlated with physical structures.
In the current implementation, we therefore only use the information
criterion approach to inform our model choices, and tailor the model
map manually according to known structures.  For instance, in the
example given in \S\,\ref{sec:planck}, we use a full four-component
model (CMB, synchrotron, free-free, and dust) inside an expanded Kp0
galactic cut \citep{bennett:2003b}, as well as in a few selected
patches (e.g., the LMC), but we ignore free-free otherwise.  However,
we expect that the information criterion approach may be developed
further, and should be a valuable tool for future experiments.

Realistically, model selection is likely to be an iterative process, as will be
demonstrated in the worked example of \S\,\ref{sec:planck}.  Typically, one
constructs an initial physically motivated model map, and performs the analysis
with that model set.  Based on the results, one then evaluates the goodness-of-fit
for each pixel (i.e., the $\chi^2$ as defined by equation \ref{eq:chisq}), and
compares the results with a simple $\chi^2$ distribution with the appropriate
number of degrees of freedom (number of frequencies minus the number of free
parameters).  If the agreement is poor, the model set can be modified, or 
offending pixels can be rejected from further analysis.

Using such methods, model selection is likely to be a relatively straightforward
(although somewhat tedious) process for future experiments.  However, in the
present paper we are satisfied with a very simple choice of models, based on
established sky masks.

\subsection{Parameter estimation by MCMC}

By now, parameter estimation by Markov Chain Monte Carlo is a well established
technique within the CMB community, with its most visible application being
estimation of cosmological parameters from the CMB power spectrum.  To our
knowledge, it has not yet been applied to component separation, and we therefore
briefly describe the algorithm here. For more details, we refer the interested
reader to, e.g., \citet{lewis:2002} or \citet{verde:2003}.

\subsubsection{Algorithm}

Suppose we want to estimate a set of parameters and corresponding uncertainties
from a set of observed data, and that we know how to compute the likelihood given
an arbitrary combination of parameters.  The Markov Chain Monte Carlo algorithm is
then given by the following simple steps:

\begin{enumerate}

\item Choose any initial point in parameter space, $\theta_0$, and
  compute the corresponding likelihood, $\mathcal{L}_{0} = \mathcal{L}(\theta_0)$.

\item Define a \emph{stochastic} function $f$ that, given parameters
   $\theta_i$, returns a new set of parameters $\theta_{i+1} = f(\theta_i)$.

\item Compute $\theta_{i+1}$ given by $f$ and the corresponding
likelihood, $\mathcal{L}_{i+1} = \mathcal{L}(\theta_{i+1})$.

\item Set $\theta_{i+1} = \theta_{i}$ (i.e., reject the proposal) with
  probability $p = 1- \min(\mathcal{L}_{i+1}/\mathcal{L}_{i}, 1)$.

\item Go to step no.\ 3, and iterate as long as necessary.

\end{enumerate}
This procedure returns a chain of parameter samples $\theta_i$, $i=1, \ldots,
N_{\textrm{samples}}$, and their multi-dimensional histogram equals the likelihood
in the limit of an infinite number of samples. 

Some intuition for the process may be gained by noticing the form of
the ``jump probability'' given in step~4 of the algorithm: If the
likelihood of the proposed point is larger than that of the old point,
we never reject the proposed point; we always move towards more likely
solutions when proposed.  However, if the likelihood is smaller, we
still accept the proposed point with probability $p =
\mathcal{L}_{i+1}/\mathcal{L}_{i}$.  This guarantees that we spend
most of the time around the peak position, but still explore less
likely points. Indeed, it may be proven that the time spent at a given
point in parameter space is proportional to the likelihood itself.

\subsubsection{Automated MCMC in practice}

In practice, there are several problems connected to MCMC parameter estimation;
usually, most of these may be identified (and often solved) by simple visual
inspection of the Markov chains.  However, since we want to analyze many thousands
of independent pixels, finding automated and yet robust solutions to the same
problems is of critical importance; a solution that works in 99\% of all cases is
not good enough. 

\para{Burn-in}
Although it is true that the initial guess may be chosen arbitrarily, and that the
chain eventually will burn in to the right solution, it is difficult to construct
a truly reliable automated prescription for when burn-in has occurred.  Also,
since computational speed is of critical importance, it is not acceptable to
spend a long time in the burn-in phase.  For both of these reasons, we choose to
initialize the chains at the maximum-likelihood point, which we find using a
standard non-linear fitting algorithm. (We have found that a sequential quadratic
programming (SQP) method works very well for this task.)

\para{Proposal function}
In step 2 of the algorithm, we must establish a proposal function $f$.  A simple
example of such a function would be $\theta_{i+1}^j = \theta_i^j + \delta \theta^j
\eta^j$, where $j$ is a parameter index, $\delta\theta^j$ is a pre-defined rms
step size, and $\eta^j$ is a Gaussian stochastic variate of zero mean and unit
variance. However, since most parameters of interest are usually strongly
correlated, this choice is quite inefficient.

Our current best solution is to run a preliminary chain (using the
univariate Gaussian proposal function described above, with manually
set step sizes), and compute the covariance matrix $C_{ij} =
\bigl<\delta\theta^i \delta\theta^j\bigr>$ of the resulting samples.
We then Cholesky-decompose this matrix, $\mathbf{C} =
\mathbf{L}\mathbf{L}^{\textrm{t}}$, and define our new proposal
function to be $\theta_{i+1} = \theta_{i} + \alpha\mathbf{L} \eta$,
where $\eta$ now is a vector of Gaussian variates and $\alpha$ is an
overall scale factor, typically initialized at $\sim0.3$.  This
ensures that the proposed samples have approximately the correct
covariance structure, and the overall sampling efficiency is thereby
greatly improved.

To avoid too large or too small step sizes, we also impose the requirements that
the acceptance ratio (the ratio of accepted to rejected proposals) must be higher
than 5\%, and lower than 80\%.  If one of these two criteria is violated, we
divide or multiply $\alpha$ by 2, and re-start the MCMC analysis.

\para{Convergence} 
Finally, we must decide when a sufficient number of samples has been accumulated.
No general solutions are available.  We have adapted a good working solution
proposed by \citet{gelman:1992}, as follows.  Run $m$ independent MCMC chains in
parallel for the same pixel, each producing $n$ samples.  Compute the
following quantities,
\begin{align}
W &= \frac{1}{m(n-1)}\sum_{j=1}^{m}\sum_{i=1}^{n} \left(\theta_{i}^{j}
- \hat{\theta}_{i}^{j}\right)^2, \\
B &= \frac{n}{m-1} \sum_{j=1}^{m}\left(\theta^j-\hat{\theta}\right)^2,
\\
V &= \left(1-\frac{1}{n}\right) W + \frac{1}{n} B,\\
R &= \frac{V}{W},
\end{align}
where $\hat{\theta}$ is the average over all $m\cdot n$ samples, and
$\hat{\theta}^j$ is the average of the samples within chain number $j$.  $W$
estimates the variance within each chain individually, while $B$ estimates the
variance between the chains.  When the chains have converged properly, $V$ and $W$
should be identical, and $R$ should be close to unity.

\citet{gelman:1992} make the general recommendations that the initial points for
the $m$ chains should be over-dispersed relative to the true distribution, and
that the chains should be run until $R<1.2$. However, as discussed above, we
initialize the chains at the maximum likelihood value in order to avoid burn-in
problems, and thus the first point is certainly not fulfilled in our approach. 
Consequently, the numerical value of the convergence criterion they give does not
apply to our prescription.

To remedy this situation, we impose two alternative criteria.  First, we require
that the chains run for a minimum number of samples (typically on the order of
$10^7$, but only storing every, say, 500th sample, to reduce correlations). 
Second, we require that the largest value of $R$, individually computed for all
parameters included in the model, must be smaller than 1.01. In most cases, we
find that the latter criterion is fulfilled long before the former, indicating
that the overall CPU time may be decreased somewhat. However, since we cannot
inspect the chains manually for more than a few pixels, we consider safety to be
more important than speed, and adopt a very conservative approach.  With the
criteria discussed here, we have found excellent convergence in all cases we have
inspected, and the Monte Carlo error (the error on the error due to a finite
number of samples) is typically less than 1\%.

\subsubsection{Point estimators and uncertainties}

The MCMC algorithm provides us with a large number of multi-variate samples drawn
from the likelihood, and these may be used to form a great variety of useful
statistics.  Here we focus on the uni-variate distributions for each parameter,
marginalizing over all others.  (Marginalization with MCMC samples is
straightforward: simply disregard the ``uninteresting'' parameters, and make a
histogram of the ``interesting'' parameter sample values. Priors are discussed
briefly in the next section.)

Our point estimate for each parameter value is then defined to be the mean of the
MCMC samples, with an uncertainty given by variance of the samples.  As shown
later in practical examples, this Gaussian approximation is quite good for both
the CMB temperature and the spectral indices, while the distributions for the
synchrotron and free-free amplitudes tend to be non-Gaussian due to a combination
of strong internal correlations and a positivity prior.

One of the most important differences between our approach and that
taken by \citet{brandt:1994} is that they chose the maximum likelihood
value as their point estimate, whereas we choose the mean.  This makes
our estimate considerably more stable with respect to noise, since it
takes more to shift the entire likelihood volume than to change its
shape.  Therefore, not only does our method yield accurate error bars
on all relevant quantities, but also the point estimates are more
reliable.

\subsubsection{Priors}

Bayes' theorem states that the posterior distribution $P(\theta|\mathbf{d})$ is
given by 
\begin{equation}
P(\theta|\mathbf{d}) \propto P(\mathbf{d}|\theta)\,P(\theta) = \mathcal{L}(\theta)\, P(\theta),
\end{equation}
where $\mathcal{L}(\theta)=P(\mathbf{d}|\theta)$ is the likelihood and
$P(\theta)$ is a prior. Implicit in the above discussion, we have
adopted the simplest possible choice for this paper, namely uniform
priors between two (not necessarily finite) limits. Specifically, we
impose no constraints on the CMB temperature, and only a positivity
constraint ($A > 0$) on the foreground amplitudes. For the spectral
indices, our priors are sufficiently generously chosen to never exclude
physically realizable values ($-4<\beta_{\textrm{s}} < -2.2$ for
synchrotron, and $1 < \beta_{\textrm{d}} < 3$ for thermal
dust). However, anticipating a future \emph{WMAP}-only analysis,
priors may then become considerably more important due to the high
noise levels of the experiment.

\subsection{Analytic estimation of linear parameters}
\label{sec:analytic_estimation}

The low-resolution MCMC analysis is by far the most expensive step in our
algorithm.  Once this step has been completed, estimation of the component
amplitudes may be done both accurately and efficiently by solving sets of linear
equations, with negligible computational resources.

Recall that our likelihood function takes the form of a standard $\chi^2$,
\begin{equation}
\chi^2 = \sum_{k = 1}^{N_{\textrm{band}}} \frac{1}{\sigma^2_k}
\left(d_{k} - \sum_{i=1}^{N_{\textrm{comp}}} A_{i} \,c_{ik}
\left(\frac{\nu_{ik}}{\nu_{i0}}\right)^{\beta_i}\right)^2.
\end{equation}
Since the only free parameters in the problem are now the component amplitudes
$A_{i}$, and the noise is assumed to be Gaussian, this is simply a multi-variate
Gaussian distribution.  Therefore the mean of the distribution equals the maximum
likelihood value, and may be determined simply by equating the derivatives of the
$\chi^2$ with respect to the parameters to zero.  In a matrix form, this reads
\begin{equation}
\frac{\partial \chi^2}{\partial \mathbf{a}} = 0 \quad \Rightarrow\quad
\mathbf{a} = \mathbf{M}^{-1}\mathbf{d},
\label{eq:linsys}
\end{equation} 
where $\mathbf{a}^{\textrm{t}} = (A_1, \ldots,
A_{N_{\textrm{comp}}})^{\textrm{t}}$,
\begin{equation}
M_{ij} = \sum_{k=1}^{N_{\textrm{band}}} \frac{c_{ik} \,c_{jk}}{\sigma^2_k} 
\left(\frac{\nu_{ik}}{\nu_{i0}}\right)^{\beta_i} 
\left(\frac{\nu_{jk}}{\nu_{j0}}\right)^{\beta_j}, 
\label{eq:m_matrix}
\end{equation}
and
\begin{equation}
d_{i} = \sum_{k=1}^{N_{\textrm{band}}} \frac{d_{k} \, c_{ik}}{\sigma^2_k} 
\left(\frac{\nu_{ik}}{\nu_{i0}}\right)^{\beta_i}.
\label{eq:d_vector}
\end{equation}

While the above formulae yield excellent results for the high-resolution parameter
point estimates (as long as all amplitudes are non-negative), reliable error
estimation within this framework is complicated.  One problem is that we need to
propagate the errors in the spectral indices into the final data products, taking
into account the strong correlations between the errors.  A second problem is
introduced by the positivity prior on the foreground amplitudes, which leads to
strongly non-Gaussian distributions when active.  Nevertheless, for well-behaved
pixels (i.e., those with clear detection of all components individually) a rough
approximation may be established by means of the usual error propagation formula.

Suppose we are interested in a quantity $z = f(x,y,\ldots)$ that depends on a set
of measured quantities $x, y, \ldots$, each with \emph{independent} and
\emph{Gaussian} errors $\Delta x, \Delta y, \ldots$. In this case, the
uncertainty $\Delta z$ is given by
\begin{equation}
\Delta z^2 = \left(\frac{\partial f}{\partial x}\right)^2 \Delta x^2 + 
\left(\frac{\partial f}{\partial y}\right)^2 \Delta y^2 + \cdots.
\end{equation}
This may be applied to our case by making the identification $A_i = f(\mathbf{x})
= \sum_{j = 1}^{N_{\textrm{comp}}} M^{-1}_{ij} d_{j}$. The uncertain quantities
are both the observed data and the non-linear parameters, $\mathbf{x}^{\textrm{t}}
= (d_{\nu}, \beta)^{\textrm{t}}$.

To compute the uncertainties, we need the partial derivatives, which by equation
\ref{eq:linsys} read
\begin{align}
\frac{\partial \mathbf{a}}{\partial \mathbf{x}} = \mathbf{M}^{-1} \frac{\partial
  \mathbf{M}}{\partial \mathbf{x}} \mathbf{M}^{-1} \mathbf{d} + \mathbf{M}^{-1}
  \frac{\partial \mathbf{d}}{\partial \mathbf{x}}.
\end{align}
The derivatives $\partial\mathbf{M}/\partial \mathbf{x}$ and $\partial
\mathbf{d}/\partial \mathbf{x}$ are obtained from equations \ref{eq:m_matrix}
and \ref{eq:d_vector}.

Great care must be taken when applying this method to the high-resolution
data---it is strictly valid only under the assumptions that the uncertain
quantities are both Gaussian distributed and internally independent, neither of
which is true for our problem.  Nevertheless, while the formal requirements are
not strictly fulfilled, the approximation may still be useful for establishing
the order of magnitude of the uncertainties.

\begin{deluxetable*}{lccccc}
\tablewidth{0pt}
\tablecaption{Frequency band specifications\label{tab:specifications}} 
\tablecomments{Specifications for each frequency band used in the
  simulation. All beams are assumed to be
  Gaussian. The RMS values for $7'$ pixels are computed taking into
  account the scanning strategy of each detector, but neglecting noise
  correlations. The RMS values per $60'$ beam are estimated from
  1000 Monte Carlo simulations by drawing Gaussian random numbers
  corresponding to the RMS level of $7'$ pixels, deconvolving
  the instrument beam, and finally convolving with a $60'$ FWHM Gaussian
  beam. The values also take into account reduced pixel resolution, from
  $7'$ to $14'$ pixels. The \emph{Planck} RMS values are requirement levels, 
  not goals. }
\tablecolumns{6}
\tablehead{ Experiment & Center frequency & Bandwidth & Beam FWHM &
  RMS per $7'$ pixel& RMS per $60'$ beam
  \\ & (GHz) & (GHz) & (arcmin) & ($\mu\textrm{K}$) &  ($\mu\textrm{K}$)}
\startdata
\emph{WMAP} &  \phn23 & \phn\phn5  & 52.8    & $50\phn\pm\phn7$    & $7.6\phn\pm\phn1.1$\\
LFI  &  \phn30 & \phn\phn6  & 33.0    & $33\phn\pm\phn7$    & $3.0\phn\pm\phn0.6$\\
\emph{WMAP} &  \phn33 & \phn\phn8  & 39.6    & $51\phn\pm\phn7$    & $5.0\phn\pm\phn0.7$\\
\emph{WMAP} &  \phn41 & \phn11     & 30.6    & $49\phn\pm\phn8$    & $4.3\phn\pm\phn0.7$\\
LFI  &  \phn44 & \phn\phn8  & 24.0    & $33\phn\pm\phn6$    & $2.7\phn\pm\phn0.5$ \\
\emph{WMAP} &  \phn61 & \phn16     & 21.0    & $60\phn\pm\phn9$    & $4.8\phn\pm\phn0.7$\\
LFI  &  \phn70 & \phn14     & 14.0    & $31\phn\pm\phn6$    & $2.4\phn\pm\phn0.5$ \\
\emph{WMAP} &  \phn94 & \phn24     & 13.2    & $73\phn\pm11$       & $5.6\phn\pm\phn0.8$\\
HFI  & 100     & \phn33     & \phn9.5 & $14\phn\pm\phn3$    & $1.1\phn\pm\phn0.2$ \\
HFI  & 143     & \phn48     & \phn7.1 & $\phn8\phn\pm\phn1$ & $0.6\phn\pm\phn0.1$ \\
HFI  & 217     & \phn72     & \phn5.0 & $11\phn\pm\phn2$    & $0.9\phn\pm\phn0.2$
\enddata

\end{deluxetable*}

First consider the assumption of independence.  The data points are
for all practical purposes uncorrelated with the non-linear
parameters, since the latter are estimated from heavily smoothed maps,
and therefore averaged over many pixels.  Additionally, if the model
only contains one spectral index, or two weakly coupled indices (e.g.,
synchrotron and dust indices, but not synchrotron and free-free), the
independence assumption is thus nearly fulfilled.

The assumption of Gaussian distributions is generally more problematic.  On the
one hand, if a particular foreground component is near or below the detection
limit, the actual distribution is sharply cut off for negative values.  On the
other hand, if two components have similar spectral dependencies (like
synchrotron and free-free emission), the corresponding parameter probability
distributions are usually strongly skewed.  In either case, the rms is not an
accurate estimator of the uncertainty.

\subsection{Propagation of errors to the CMB power spectrum and cosmological parameters}
\label{sec:gibbs}

The analytical approach described in the previous section yields good point
estimates for the desired parameters, but only approximate uncertainties.
Further, it is not straightforward to propagate the errors further into
higher-level data products such as the CMB power spectrum or cosmological
parameters. 

A much more powerful solution may be devised by combining the methods described in
the present paper with the Gibbs sampling approach of \citet{jewell:2004},
\citet{wandelt:2004} and \citet{eriksen:2004b}.  Whereas most other techniques
only provide the user with a very simple description of the power spectrum
probability distribution (e.g., a maximum-likelihood estimate and a Fisher
matrix), the Gibbs sampling approach yields the complete multi-variate
probability density $P(C_{\ell}|\mathbf{d})$, $C_{\ell}$ being the CMB power
spectrum and $\mathbf{d}$ the data.  Further, it is very straightforward to
introduce new sources of uncertainty into the framework, and such uncertainties
are then seamlessly propagated through to the final data products.  We
outline here how foreground uncertainties may be propagated to the CMB power
spectrum and cosmological parameters, but leave the details for a future
publication. 

\begin{figure}

\mbox{\epsfig{file=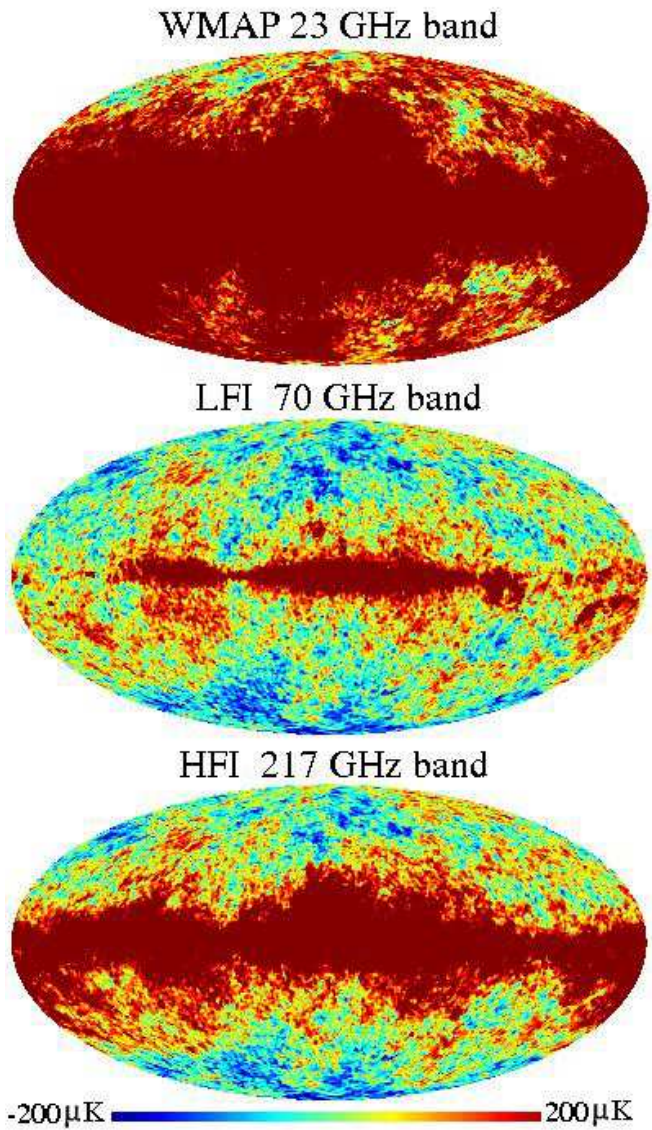,width=\linewidth,clip=}}

\caption{The ``high-resolution'' simulations used in this paper.  Shown
  are the 23\,GHz channel from the \emph{WMAP} experiment, the 70\,GHz
  channel from the LFI experiment, and the 217\,GHz channel from the
  HFI experiment.  All maps are smoothed to a common resolution of
  $1^\circ$ FWHM.}
\label{fig:sim_bands}
\end{figure}

The Gibbs sampling approach is similar in philosophy to the MCMC method that is
used extensively in this paper: the target density is established by drawing
samples from it.  In our case, we are interested in the joint probability
distribution $P(C_{\ell},\mathbf{s}_{\textrm{CMB}}, \mathbf{s}_{\textrm{s}},
\mathbf{s}_{\textrm{ff}}, \mathbf{s}_{\textrm{d}}|\mathbf{d})$, where
$\mathbf{s}_i$ are the four interesting signal components discussed earlier, CMB,
synchrotron, free-free, and thermal dust emission.  While it is difficult to
sample from this distribution directly, the Gibbs sampling algorithm provides a
neat solution.  Suppose we want to draw samples from a joint distribution
$P(x,y)$, but only know how to sample from the conditional densities $P(x|y)$ and
$P(y|x)$.  In that case, the theory of Gibbs sampling says that samples $(x,y)$
can be drawn by iterating the following sampling equations,
\begin{align}
x^{i+1} &\leftarrow P(x|y^{i}), \\
y^{i+1} &\leftarrow P(y|x^{i+1}).
\end{align}
The symbol '$\leftarrow$' indicates that a random number is drawn from the
distribution on the right hand side. After some burn-in period, the samples will
converge to being drawn from the required joint distribution. 

Suppose now that we want to analyze a data set that for simplicity only includes
CMB and synchrotron emission, the latter being parameterized by an amplitude
$A_{\textrm{s}}$ and a spectral index $\beta_{\textrm{s}}$ for each pixel. 
Suppose further that we already have run an MCMC analysis for each pixel as
described earlier, and have access to the corresponding probability distributions.
In that case, the Gibbs sampling algorithm may be applied by means of the
following sampling chain:
\begin{align}
\beta_{\textrm{s}}^{i+1} &\leftarrow
  P(\beta_{\textrm{s}}|C_{\ell}^{i}, A_{\textrm{s}}^{i}, \mathbf{s}_{\textrm{CMB}}^{i},
  \mathbf{d}),\\
A_{\textrm{s}}^{i+1} &\leftarrow
  P(A_{\textrm{s}}|C_{\ell}^{i}, \beta_{\textrm{s}}^{i+1}, \mathbf{s}_{\textrm{CMB}}^{i},
  \mathbf{d}),\\
\mathbf{s}_{\textrm{CMB}}^{i+1} &\leftarrow
  P(\mathbf{s}_{\textrm{CMB}}|C_{\ell}^{i}, A_{\textrm{s}}^{i+1},
  \beta_{\textrm{s}}^{i+1}, \mathbf{d}),\\
C_{\ell}^{i+1}     &\leftarrow P(C_{\ell}|\mathbf{s}_{\textrm{CMB}}^{i+1}).
\end{align}
(The CMB power spectrum $C_{\ell}$ only depends on the CMB signal, not the
foregrounds, and therefore the other components are omitted from the right-hand
side in the last equation.)  The first two rows are to be performed for each pixel
individually, while the last two rows are performed in harmonic space, reflecting
the intuitively pleasing idea that foregrounds should be handled in pixel space,
while CMB fluctuations are better handled in harmonic space. 

To perform the analysis as described above, we have to be able to
sample from all involved conditional distributions.  Sampling the CMB
signal and power spectrum parts is detailed by, e.g.,
\citet{eriksen:2004b}.  Sampling the foreground amplitudes given the
spectral indices is straightforward, since the corresponding
distributions are simple Gaussians.

However, sampling the spectral indices is \emph{a-priori} not trivial---their
distributions are highly non-Gaussian, and no analytical expressions exist. 
However, given that we already have run a MCMC analysis whose product is precisely
the joint density $P(\mathbf{s}_{\textrm{CMB}}, A_{\textrm{d}},
\beta_{\textrm{s}}|\mathbf{d})$, the problem is mostly solved.  We could simply
generate a full multi-dimensional histogram from the MCMC samples (for each pixel
separately), and pick out the synchrotron index column that corresponds to the
other currently fixed parameter values.  Given this one-dimensional distribution,
we could then sample numerically using standard techniques. 

For completeness, we note that adding more than one foreground component is a
trivial extension of this scheme.  Component amplitudes are added individually,
while there is a choice for spectral indices---one may either sample these
individually, as done above for the synchrotron index, or for greater efficiency,
one may also exploit the multi-variate information given by the MCMC analysis.

With the above prescription, it is finally possible to propagate the foreground
uncertainties rigorously all the way from the observed data through to the CMB
power spectrum, and therefore to the cosmological parameters.  Further, with this
approach one also obtains full-resolution sampled uncertainties of the component
amplitudes, as opposed to the analytical approximations discussed in the previous
section, and a complete probabilistic description of the system is thereby
established.

In the form described above, the algorithm requires storage of a multi-dimensional
histogram for each pixel.  This is tremendously wasteful, and clearly not feasible
due to memory limitations.  A first necessary step is therefore to compress the
likelihood information into a smaller set of numbers.  One possible solution was
proposed by \citet{verde:2003}, who fitted a 4th order polynomial to the
log-likelihood,
\begin{equation}
\begin{split}
\ln \mathcal{L} = q_0 + \sum_{i} q_1^i \delta_i + 
\sum_{ij} q_2^{ij} \delta_i \delta_j +\\
+ \sum_{ijk} q_3^{ijk} \delta_i \delta_j \delta_k +
\sum_{ijkl} q_4^{ijkl} \delta_i \delta_j \delta_k \delta_l.
\end{split}
\label{eq:lnl_compress}
\end{equation}
Here $q_i$ are fit coefficients, and
$\delta_i=(\alpha_i-\alpha_i^0)/\alpha_i$, $\alpha_i$ being parameter
number $i$ and $\alpha_i^0$ its maximum likelihood
value\footnote{Introducing normal parameters and/or higher-order
expansions may be required for obtaining sufficient accuracy; see
e.g., Chu, Kaplinghat \& Knox (2003) for an application to
cosmological parameters.}.  With only 210 coefficients for a
six-parameter likelihood, this approximation at least satisfies the
memory requirements for practical usage in our problem.  Further, it
is straightforward to map out the required spectral index
distributions given a set of component amplitudes using this form.
Therefore, considering that the process parallelizes trivially, there
seem to be no unsurmountable computational problems connected to this
approach.

While Gibbs sampling as currently implemented has problems with
probing the low signal-to-noise ratio regime properly, it works very
well for signal-to-noise ratios larger than unity
\citep{eriksen:2004b}, and this is exactly where the foreground
uncertainties dominate.  Therefore it seems reasonable to use the
approach presented here to analyze the high and intermediate
signal-to-noise ratio regimes, propagating foreground uncertainties to
the final products, and a standard ``Master''-type analysis
\citep{hivon:2002} for the low signal-to-noise regime, at the cost of
neglecting foreground uncertainties at these angular scales.

\section{Example: Application to \emph{Planck} and six-year \emph{WMAP} data}
\label{sec:planck}

We now apply the MCMC component separation method described in
\S\,\ref{sec:algorithm} to simulations of the current \emph{WMAP} and the future
\emph{Planck} missions.  We first give a detailed presentation of the simulations
and data.  We then study the behavior of the algorithm for one arbitrarily chosen
pixel, before considering the full sky map solutions. 

We point out that our main goal in this paper is to study the algorithm itself,
and not to simulate an actual data release.  We therefore choose examples both
with and without modeling errors, in order to illustrate problems that may be
encountered in an analysis of real data.

\subsection{Simulations and models}
\label{sec:fgs}

The simulations used in the following are constructed as a sum of a cosmological
CMB signal, three foreground components (synchrotron, free-free, and thermal dust
emission) and instrumental noise.  We include five bands (centered at 23, 33, 41,
61 and 94\,GHz) from \emph{WMAP}, three bands (30, 44 and 70\,GHz) from the
\emph{Planck} Low Frequency Instrument (LFI), and three bands (100, 143 and
217\,GHz) from the High Frequency Instrument (HFI), for a total of eleven bands
between 23 and 217\,GHz\footnote{The three highest HFI frequency bands are not
included in the analysis because they would introduce significant dust modeling
errors; we simulate dust with a two-component model, but fit for a one-component
model.}.  Specifications for each detector are given in
Table~\ref{tab:specifications}.

\para{CMB}The CMB component is assumed to be Gaussian distributed, with variances
given by the best-fit \emph{WMAP} power-law power spectrum \citep{bennett:2003a,
hinshaw:2003, spergel:2003}, including multipoles between $\ell=2$ and 1024.  The
signal realization is filtered through the HEALPix pixel window function and the
instrument specific beam windows.  (Since all the \emph{Planck} beam windows are
not available, we choose for simplicity to model even the \emph{WMAP} beams as
Gaussians with appropriate FWHM's.)

\para{Noise}The noise is assumed to be Gaussian and uncorrelated, but non-uniform
according to the scanning strategy of each detector.  For \emph{WMAP}, we assume
a six-year mission, and rescale the published first-year sensitivity levels by
$1/\sqrt{6}$. For \emph{Planck}, we adopt the requirement levels, which are a
factor of two worse than the goals, for the baseline one-year mission.

\begin{figure*}

\mbox{\epsfig{file=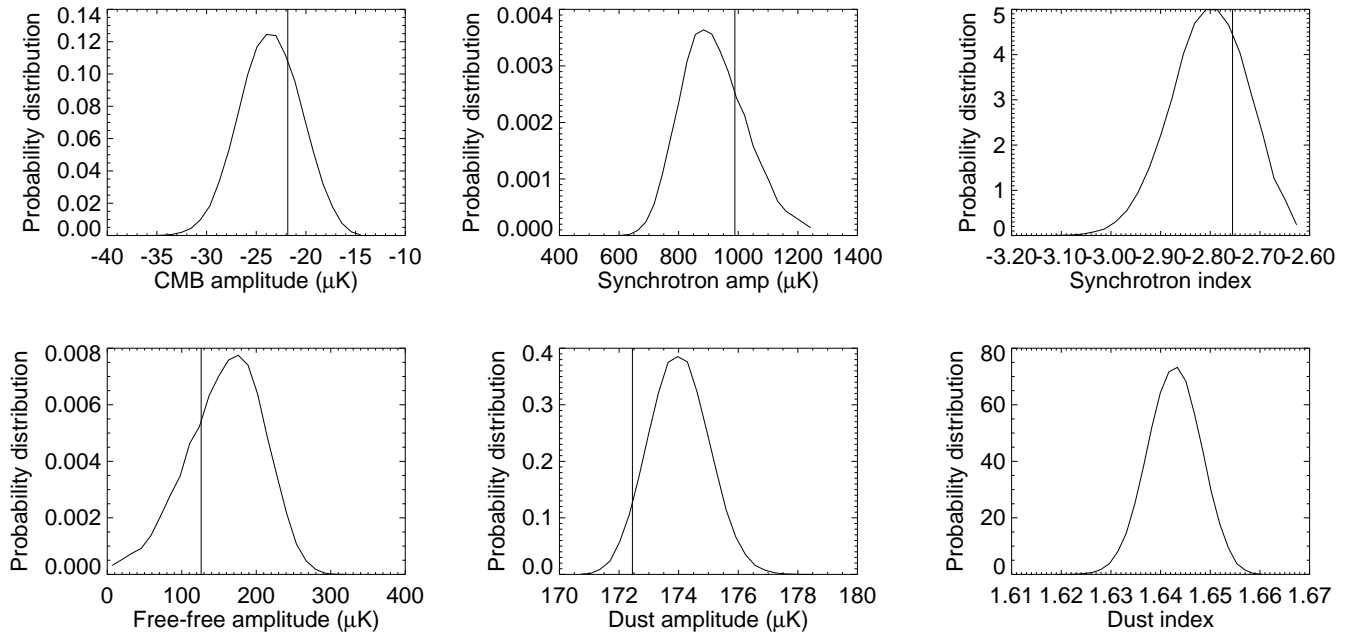,width=\linewidth,clip=}}

\caption{Marginalized parameter probability distributions for an
  arbitrarily chosen pixel inside the Galactic plane, generated by
  MCMC as described in the text.  The vertical lines show the true
  input value for the pixel.  (The true value is not well defined for
  the dust spectral index, since dust is modeled by a two-component
  spectrum, while a one-component model is fitted.  However,
  the stronger of the two dust components has an index parameter of
  $\alpha_1 = 1.67$.)}
\label{fig:marg_1d_dist}
\end{figure*}

\para{Synchrotron emission}The only all-sky map currently available to provide a
template of Galactic synchrotron emission in HEALPix format is the 408\,MHz
survey by \citet{haslam:1982}.  This has a resolution  of only 51\arcmin, thus
additional power must be added on smaller angular scales for our purposes in this
paper.  We adopt the model of \citet{giardino:2002}, who estimated the amplitude
and slope of the synchrotron angular power spectrum at low Galactic latitudes for
$l\ge 150$.  A Gaussian realization was then generated from that power spectrum to
which was applied a Galactic modulation, multiplying the signal in each pixel by
the ratio between the \citet{haslam:1982} template in that pixel and the
maximum.  The final template is added to the original \citet{haslam:1982} map. 
An all-sky template for the synchrotron spectral index was estimated combining
the all-sky data from \citet{haslam:1982}, with the Northern sky observations of
\citet{reich:1986} at 1420\,MHz and the southern sky counterpart of
\citet{jonas:1998} at 2300\,MHz\footnote{This hybrid spectral index model should
ultimately be superseded by the full-sky 1420\,MHz survey described in 
\citet{reich:2003}.}. This constitutes our synchrotron model.

\begin{figure*}

\mbox{\epsfig{file=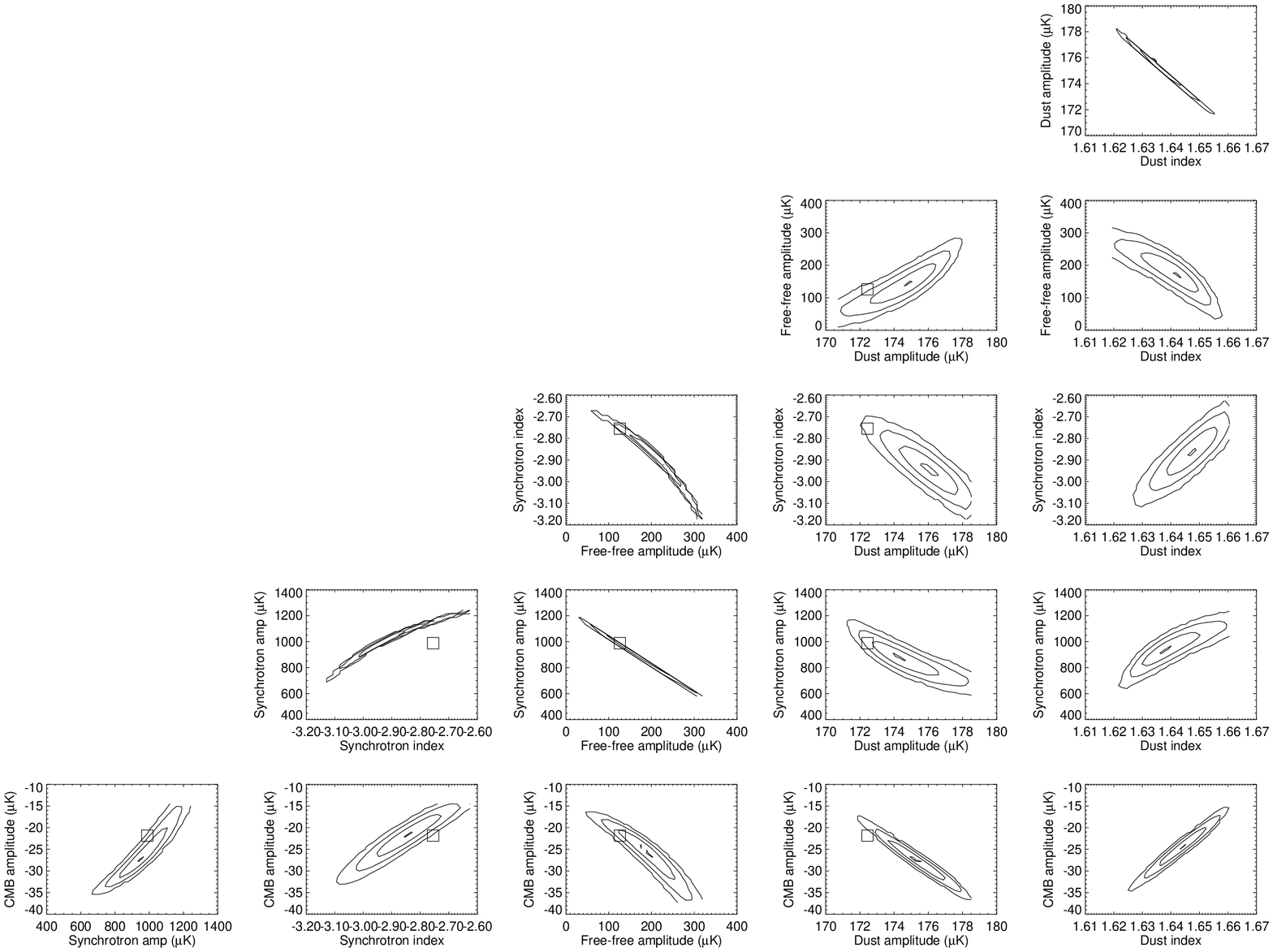,width=\linewidth,clip=}}

\caption{Marginalized two-dimensional probability distributions for the same
pixel as in Figure~\ref{fig:marg_1d_dist}, computed by \hbox{MCMC}.  Boxes
indicate the true input values, and the contours mark the peak and the 68, 95 and
99.7\% confidence levels.}
\label{fig:marg_2d_dist}
\end{figure*}

\para{Free-free emission}%
Current models of the free-free emission exploit the expected correlation with
H$\alpha$ emission (see Dickinson et al.~2003 and references therein), and
predict free-free emission (given in antenna temperature) as follows
\begin{equation}
\begin{split}
s_{\textrm{ff}}(\nu)= A_{{\rm H}\alpha} \cdot 0.1366\cdot10^{0.029\cdot\frac{10^4{\rm
K}}{T_{\textrm{e}}}} \,G \,\cdot\quad\\
\cdot\left(\frac{T_{\textrm{e}}}{10^{4}{\rm
K}}\right)^{0.517} \left(\frac{\nu}{\nu_{\textrm{ref}}}\right)^{-2}.
\end{split}
\label{eq:ha_freefree_scaling}
\end{equation}
Here $A_{\textrm{H}\alpha}$ is the H$\alpha$ amplitude,
$G=3.96\,T_{\textrm{e}}^{0.21}(\nu /\nu_{0,\textrm{ff}})^{-0.14}$ is the Gaunt
factor \citep{finkbeiner:2003}, $T_{\textrm{e}}$ is the electron temperature, and
$\nu_{0,\textrm{ff}}=40\,\textrm{GHz}$ is a reference frequency.

A major uncertainty when using H$\alpha$ as a template is due to the absorption
of H$\alpha$ by foreground dust, but this can be estimated using the $100\,\mu$m
maps from \citet{sfd98}.  The largest uncertainty, however, is related to the
fraction of dust ($f_{d}$) lying in front of the H$\alpha$-emitting region. 
\citet{dickinson:2003} show that for regions $30^\circ < l < 60^\circ$ and
$5^\circ < |b| < 15^\circ$, $f_{d}\approx 0.3$, while for local high latitude
regions such as Orion and the Gum Nebula, there is little or no absorption by
dust ($f_{d} \sim 0$).  The latter is supported by the cross-correlation analysis
of \citet{banday:2003} of the H$\alpha$ data with \emph{COBE}-DMR, which
contrasts with the value of $f_{d} \sim 0.5$ expected under the assumption that
the ionized gas and dust are coextensive along the line of sight (i.e., uniformly
mixed), as is assumed in the \emph{WMAP} analysis of \citet{bennett:2003b} and
\citet{fink:2004a}.

We correct for dust absorption, by assuming a single component dust
model, with a temperature of 18.3\,K, and a absorption fraction of
0.33 up to a flux corresponding to 1~magnitude.  We assume an electron
temperature of $T_{\textrm{e}}=7000$\,K, and therefore an effective
frequency scaling close to $\beta^{-2.14}$ over the range of
frequencies considered here. For the future, a more accurate model,
accounting for the steepening spectral index, could be implemented as
a correction to the simple power-law model.

\para{Thermal dust}%
We adopt model 8 of \citet{fink:1999} for thermal dust emission, with parameters
$f_{1}=0.0363$, $q_{1}/q_{2}=13$, $\nu_{0,\textrm{d}}=3000$\,GHz,
$\alpha_{1}=1.67$, $\alpha_{2}=2.70$,
$T_{1}=9.4$\,K, and $T_{2}=16.2$\,K (see equations
\ref{dust_scaling}--\ref{dust_scaling_function}).  However, this is too many 
parameters to fit individually, and we therefore adopt a simpler model for
reconstruction (see equation \ref{eq:dust}).  Modeling errors of the sort to be
expected with real data will result.

\para{Data processing}All simulations are initially made at a pixel resolution of
$N_{\textrm{side}}=512$, corresponding to a pixel size of $7'$. However, since our
method requires identical beam sizes for all frequency bands, we downgrade each
band separately to $1^\circ$ FWHM (determined by the $52.8'$ FWHM beam of the
23\,GHz \emph{WMAP} band) and reduce the pixel resolution to
$N_{\textrm{side}}=256$ (by deconvolving the original beam and pixel windows, and
convolving the common $1^\circ$ FWHM beam and lower resolution pixel window).

By downgrading the data, the noise specifications are also modified. To estimate
the effective noise levels after degradation, we therefore generate 1000 noise
realizations for each band, and downgrade these in the same manner as the actual
data maps. The effective noise levels of the downgraded maps are then estimated by
taking the standard deviation of the 1000 realizations.

The data set described above constitutes our main simulation, and is
referred to in the following as ``high-resolution data''.  Examples
are shown in Figure~\ref{fig:sim_bands}.  For the non-linear parameter
estimation step using MCMC, we noise levels must be lower, as
discussed earlier.  Therefore we smooth all maps with an additional
$6^\circ$ FWHM Gaussian beam, and downgrade the pixel resolution to
$N_{\textrm{side}}=32$.  (This smoothing scale is not optimized, but
it is sufficient for the purposes of the present paper.)  Again, the
effective noise levels are determined by Monte Carlo simulations.
This smoothed data set is referred to as ``low-resolution data''.

\para{Initial model map}The initial model map is based on the \emph{WMAP} Kp0
mask \citep{bennett:2003b}.  First the excluded region of the original mask is
expanded by $10^\circ$ in all directions.  Then all accepted pixels (i.e., the
high-latitude region) are assigned the model that includes CMB, synchrotron, and
dust (both with free amplitude and spectral index), while the model for the
rejected pixels additionally includes a free-free amplitude.

\subsection{Results}

We now apply the method of \S\,\ref{sec:algorithm} to the
simulated data set described above.

\subsubsection{Single pixel results}

We first examine the performance of the MCMC algorithm by studying one single
pixel in the low-resolution data set, namely pixel number 6100, which is located
inside the Galactic plane at $l=58^\circ, b=0^\circ$.  The reasons for choosing
this pixel (or one like it) are twofold.  First, the model for this pixel
includes all three foreground components, and has thus a complicated probability
structure.  Second, the model is rejected by the goodness-of-fit test, and this
example therefore illustrates the modeling error problem.

As discussed earlier, the MCMC algorithm basically performs a random walk on the
likelihood surface, producing a set of samples from which the likelihood may be
estimated by constructing single or multi-dimensional histograms.  Examples of
such histograms are shown in Figures \ref{fig:marg_1d_dist} and
\ref{fig:marg_2d_dist}.

The first figure shows the probability distributions for each of the six included
parameters, marginalized over all other parameters.  Comparing with the true
input values (vertical lines), we see that the algorithm reproduces the correct
values, and also that the uncertainties are reasonable compared to the true
errors.

Second, in Figure~\ref{fig:marg_2d_dist} we show two-dimensional probability
distributions for the same parameters.  The true values are marked by a box.
Several points are worth noticing in this figure.  First, all parameters are
clearly correlated, and some specific pairs very tightly so.  Examples of the
latter are dust amplitude versus dust spectral index, and synchrotron amplitude
versus free-free amplitude.

Second, many of the distributions are clearly non-Gaussian, and it is clear that
a Gaussian approximation at this stage will not yield reliable errors. Still, the
structures appear to be reasonably well behaved, and in principle it may be
possible to find analytical parameter transformations that could ease the
computational burden.

Third, while most of the true values lie inside the $3\sigma$ confidence regions,
in one case, namely the synchrotron amplitude versus synchrotron index, it lies
far outside the acceptable region.  Another perspective of this is provided by the
$\chi^2$, which for this pixel is 44.  With five degrees of freedom (eleven
frequencies and six free parameters), this particular model is thus ruled out at
the 99.9999\% confidence level.  This is because we fit for a simpler model than
the one used in the simulation: the data are smoothed by a wide $6^\circ$ beam,
and the thermal dust is fitted with a one-component model, whereas the simulation
was based on a two-component model.  This may also be seen
Figure~\ref{fig:fitting_illustration}, where the fitted spectra for each component
for this pixel are plotted.  At low frequencies, the data points lie
systematically above the fitted model, resulting in a clear rejection.

However, even though the model is strongly rejected by the goodness-of-fit test,
it is important to note all of the univariate distributions are still reasonable,
and the CMB reconstruction is still useful.  Therefore, a high $\chi^2$ does not
necessarily imply that the pixel has to be discarded from further analysis, but
rather that extra care has to be taken.  Preferentially, the extra information
indicated by the high $\chi^2$ should be used to improve the model.

\subsubsection{Low-resolution full-sky maps}

\begin{figure*}

\begin{center}
\mbox{\epsfig{file=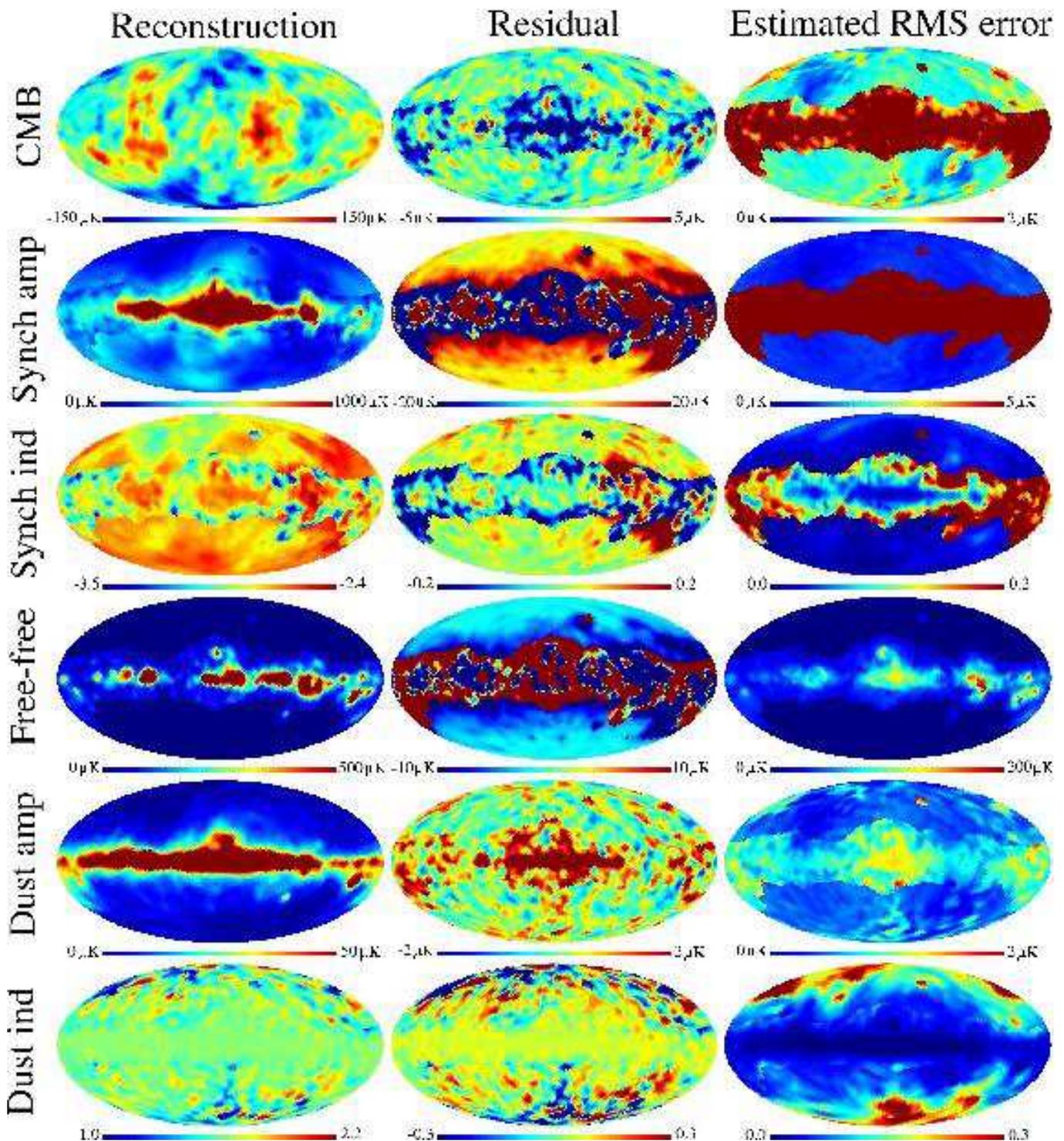,width=\linewidth,clip=}}
\end{center}

\caption{Low-resolution parameter maps reconstructed by \hbox{MCMC}.  The left
  column shows the parameter estimates, the middle column shows the
  difference between reconstructed and input maps, and the right
  column shows the rms errors estimated by \hbox{MCMC}.  From top to bottom we
  show: 1)~the thermodynamic CMB temperature; 2)~the synchrotron
  emission amplitude relative to 23\,GHz; 3)~the synchrotron spectral
  index; 4)~the free-free emission amplitude relative to 33\,GHz; 5)~the
thermal dust emission amplitude relative to 90\,GHz; and 6)~the
  thermal dust spectral index.}
\label{fig:reconstructed_maps}
\end{figure*}

\begin{figure*}

\mbox{\epsfig{file=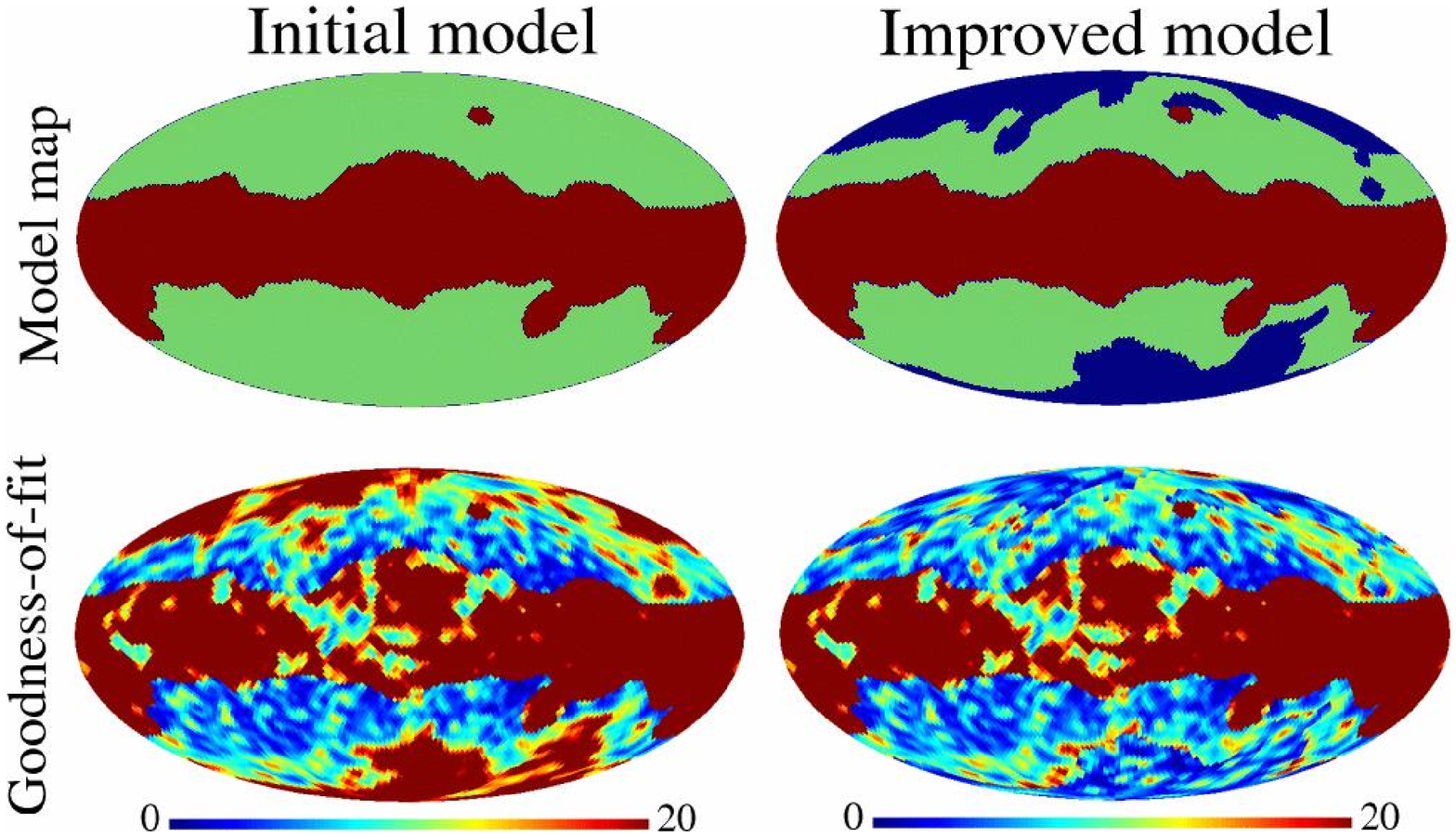,width=\linewidth,clip=}}

\caption{Example of iterative improvement of the model map. Two MCMC
  runs using the same data were made.  The first run included CMB,
  synchrotron, and dust (both with free amplitude and spectral index)
  at high latitudes (green), and also free-free at low latitudes
  (red).  The second run fixed the dust index at 1.55 at very high
  latitudes (blue).  The bottom row shows the $\chi^2$ distribution in
  the two cases; notice the significant improvement at high latitudes
  resulting from \emph{removing} a (non-critical) parameter from the
  system.  Similar improvements at low latitudes could be made by
  trial-and-error.}
\label{fig:gof_maps}
\end{figure*}

We now consider the reconstructed full-sky maps, starting with the low-resolution
maps as computed by the MCMC analysis.  The individual component maps are shown in
Figure~\ref{fig:reconstructed_maps}, with reconstructions given in the left
column, differences between reconstructed and input maps in the middle column,
and estimated errors in the right column.  In the left column of
Figure~\ref{fig:gof_maps}, we show the model map used in the analysis and the
resulting goodness-of-fit $\chi^2$ distribution (as computed with parameter
\emph{means}, not the maximum likelihood values).

Starting with the goodness-of-fit map, we first note that we should expect
$\chi^2 \lesssim 13$ at $2\sigma$ confidence at high latitudes, since the model
has six degrees of freedom in this region.  This is indeed the case for two wide
bands on each side of the Galactic plane, and both the model and the estimated
parameters may therefore be accepted as they stand.  However, at very high 
latitudes and, less surprisingly, at low latitudes, the goodness-of-fit is poor.

In the left column of Figure~\ref{fig:reconstructed_maps} we show the six
reconstructed parameter maps, and in the middle column the actual output versus
input errors.  Clearly, the method works very well, as the CMB sky map is
virtually free of artifacts, with residuals less than 10\% even in the inner
Galactic plane.  And with the exception of sharp boundaries in the foreground
reconstruction, due to different models used in different regions, the foreground
results also look encouraging. 

However, as good as these results are, we warn the reader against interpreting
them as an expected performance level for future missions.  Even though our
simulations are as realistic as possible given our current understanding of
foreground properties, they are certainly not as complicated as the real sky.
Considerable modeling errors must be expected for real data sets, and sky cuts
are very likely still required for future work.

One note about the sharp boundaries seen in the foreground maps is in order.  If
the reconstructed maps are intended for foreground studies, such features are
clearly not acceptable.  In such cases, post-processing may be required, for
instance by smoothing the boundary by a Gaussian beam.  On the other hand, if the
maps are to be used for CMB power spectrum or cosmological parameter estimation,
it is better to use the maps as they are, and propagate the pixel errors
reliably; the boundaries are mainly due to different noise properties in the
various regions.  However, we point out that the distinct boundaries seen in
Figure~\ref{fig:reconstructed_maps} are at least partially due to a poorly chosen
model map, constructed from a \emph{WMAP} galactic mask, rather than the specific
simulation under consideration; manual tweaking would surely improve the results
considerably.

Returning for a moment to the goodness-of-fit map shown in
Figure~\ref{fig:gof_maps} and comparing with the rms maps shown in the right
column of figure \ref{fig:reconstructed_maps}, we see that the very high-latitude
region with high $\chi^2$ corresponds directly to the thermal dust spectral
index map.  Further, we also see that the dust amplitude is very low in the same
region.  The interpretation is clear: thermal dust is not well constrained in
these regions because of its low amplitude, leading to poorly constrained
spectral indices.  This again propagates into the CMB component, and the overall
fit is unacceptable. 

\begin{figure*}

\mbox{\subfigure[Low-resolution MCMC results]{\label{fig:error_acc_low}\epsfig{figure=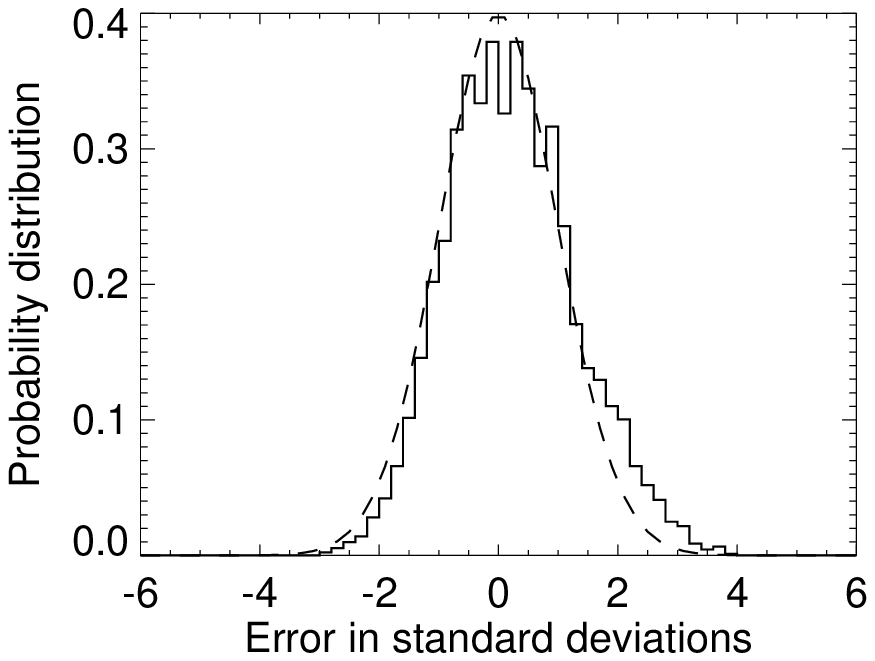,width=0.48\textwidth,clip=}}
      \quad
      \subfigure[High-resolution analytic results]{\label{fig:error_acc_high}\epsfig{figure=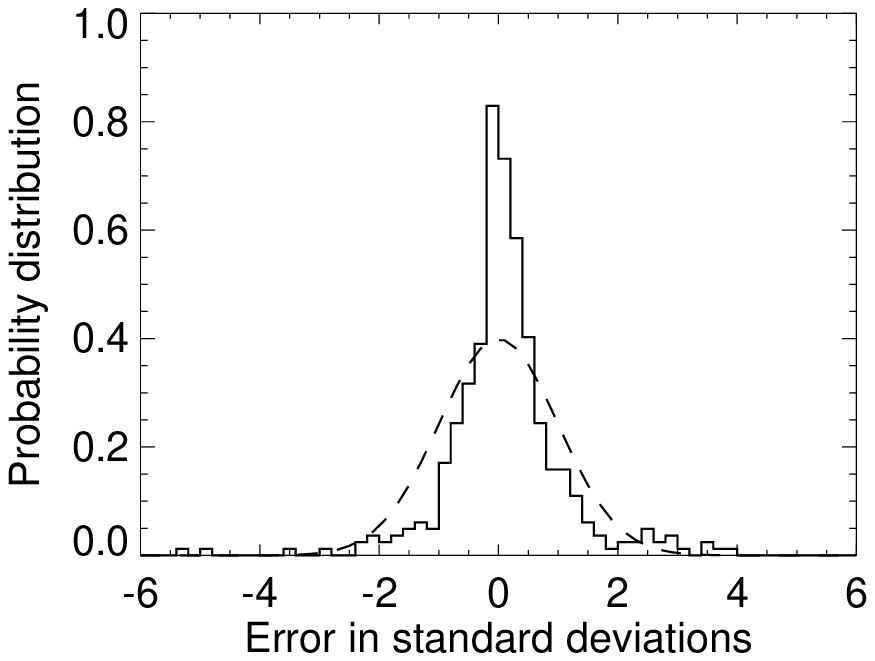,width=0.48\textwidth,clip=}}}

\caption{CMB reconstruction accuracy shown as a histogram of $\delta =
  (\Delta T_{\textrm{est}}-\Delta
  T_{\textrm{in}})/\sigma_{\textrm{est}}$ for pixels in the
  intermediate-latitude region.  Results from the low-resolution MCMC
  analysis are shown in the left panel, and from the high-resolution
  analytical analysis in the right column.  For perfect reconstruction
  of both CMB amplitude and error, both curves would match a Gaussian
  distribution with vanishing mean and unit variance (dashed curve). }
\label{fig:error_accuracy}
\end{figure*}

The solution to this problem seems obvious.  Since the main problem is
unconstrained dust spectral indices, we should manually fix them at
some reasonable value.  The potential bias introduced in the CMB and
other components by doing so is very small because of the small dust
amplitude found by the first analysis.  We implement this by assigning
a new model that fixes the dust spectral for all pixels with a dust
spectral index rms larger than 0.15 in the lower right panel of
Figure~\ref{fig:reconstructed_maps}.  The fixed spectral index value
chosen is somewhat arbitrarily chosen to be 1.55.  The modified model
map is shown in the top right panel of Figure~\ref{fig:gof_maps}.

We now repeat the analysis, and obtain the goodness-of-fit map shown in the lower
right panel of Figure~\ref{fig:gof_maps}. Clearly, introducing a new model at
high latitudes had a very beneficial impact on the results.  In principle, we
could now proceed with similar considerations at low latitudes, and obtain
reasonable fits over the full sky.  However, since our main purpose in this paper
is to illustrate the method, we are content with the slightly revised model map
shown in the top right panel of Figure~\ref{fig:gof_maps}, and use this map in
the rest of the paper.

\begin{figure}

\plotone{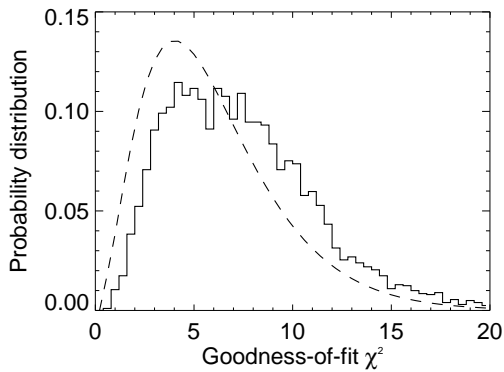}

\caption{Distribution of $\chi^2$ from the low-resolution MCMC
  analysis, shown for the intermediate latitude region, with six
  degrees of freedom. The dashed curve shows the expected 
  $\chi^{2}_{6}$ distribution.}
\label{fig:chisq_distrib}
\end{figure}

We now consider the error estimation accuracy of the MCMC algorithm.  In the left
panel of Figure~\ref{fig:error_accuracy} we plot a histogram of the relative CMB
reconstruction error $\delta= (\Delta T_{\textrm{est}}-\Delta
T_{\textrm{in}})/\sigma_{\textrm{est}}$, where $\Delta T_{\textrm{est}}$ is the
estimated CMB temperature, $\Delta T_{\textrm{in}}$ is the true value, and
$\sigma_{\textrm{est}}$ is the estimated error.  If both the amplitude and the
error are perfectly estimated, the pixel histogram will match a Gaussian
distribution with vanishing mean and unit variance.  (In this plot, we include
only pixels in the intermediate latitude region with a goodness-of-fit
$\chi^2 < 13$.)  Obviously, the algorithm works very well, as the bias is very
small indeed and the estimated error is very close to the true error.

Finally, in Figure~\ref{fig:chisq_distrib} we plot a histogram of the $\chi^2$s
of the same pixels, and compare it to a $\chi^2_6$ distribution.  Clearly, there
is a small shift towards high values, indicating that there are pixels within the
set for which the model is rejected.  Once again, for an analysis of real data, we
should then go back to the sky maps shown in Figures \ref{fig:reconstructed_maps}
and \ref{fig:gof_maps}, and try to locate the ``offending'' pixels.

We conclude this section by making a few comments on the computational cost of
the method.  Running the MCMC analysis for each pixel is by far the most
expensive step of the algorithm.  For well-behaved pixels, we find that it takes
on the order of 100~CPU seconds (divided over four processors per pixel) to reach
the convergence criteria described above.  For an $N_{\textrm{side}}=32$ map with
12,288 pixels, it therefore takes about 350~CPU hours per run.  For clusters
with of order $10^2$ processors, this is not a major problem. Further, since the
algorithm scales with the number of pixels, and parallelizes trivially, it is not
unreasonable to apply it at higher resolutions, say, at $N_{\textrm{side}}=128$
for 6000~CPU hours.

\subsubsection{High-resolution full-sky maps}

\begin{figure*}

\begin{center}
\mbox{\epsfig{file=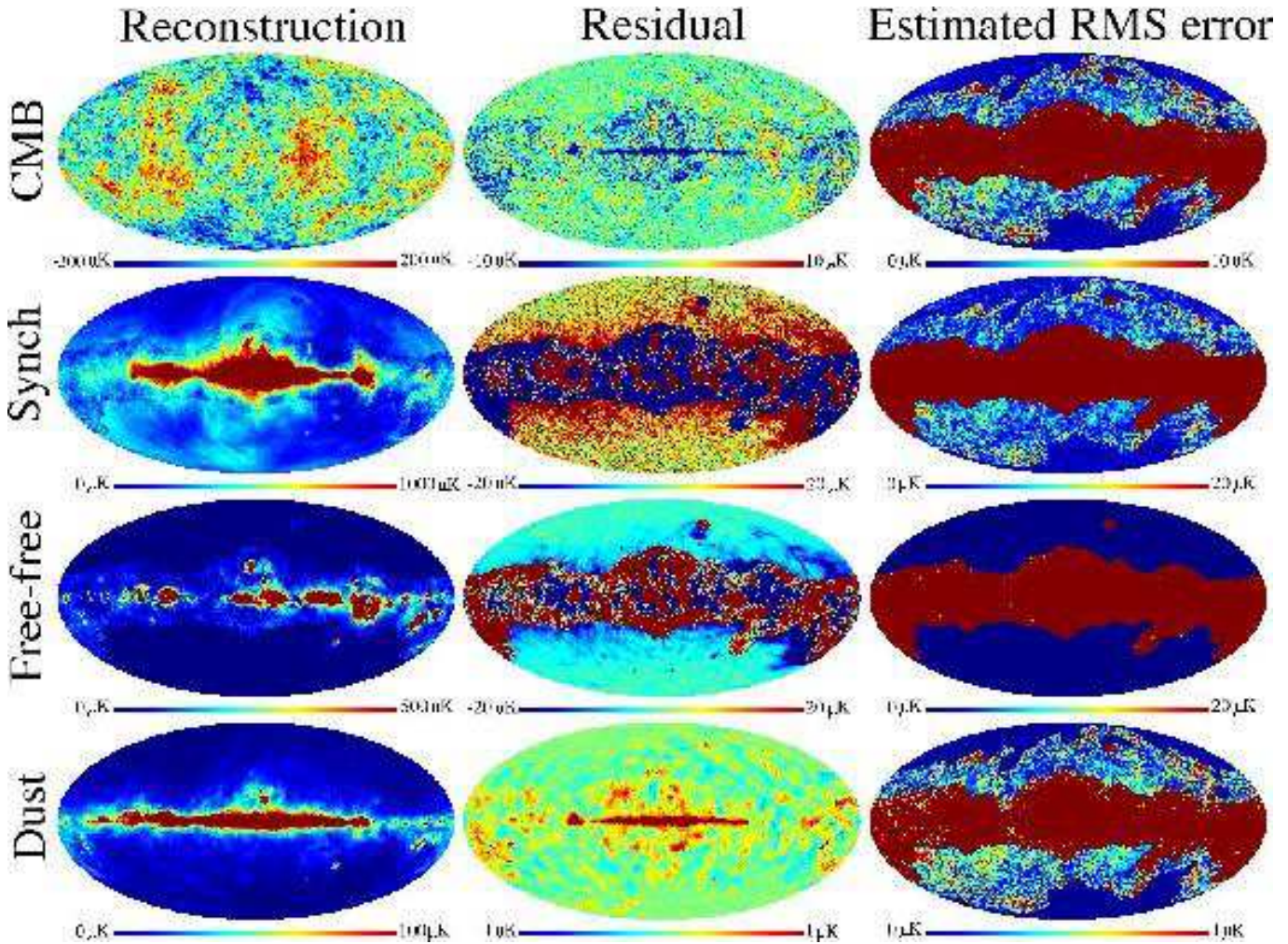,width=\linewidth,clip=}}
\end{center}

\caption{High-resolution parameter maps reconstructed by direct
  solution of linear systems, fixing spectral indices at
  low-resolution map values.  The left column shows the parameter
  estimates, the middle column shows the difference between output
  and input maps, and the right column shows the analytically estimated
  errors.  From top to bottom we show: 1)~the thermodynamic CMB
  temperature; 2)~the synchrotron emission relative to 23\,GHz; 3)~the free-free
  emission relative to 33\,GHz; and 4)~the thermal dust emission relative to
  90\,GHz.}
\label{fig:high_res_maps}
\end{figure*}

Having estimated the non-linear parameters by MCMC, the next step is to estimate
the component amplitudes from the full-resolution sky maps.  As discussed
earlier, this can be done either with a Gibbs sampling approach or with an
analytic approach.  In this paper, we choose the latter route, and leave the
former to a future publication.

The results from applying the method described in
\S\,\ref{sec:analytic_estimation} are shown in Figure~\ref{fig:high_res_maps}.
Once again, we see that the reconstructed parameter maps look visually
compelling.  There are few visible signs of contamination in the CMB
reconstruction, and, indeed, even inside the central Galactic plane the errors
are only a few tens of $\mu\textrm{K}$.

\begin{figure}

\begin{center}
\mbox{\epsfig{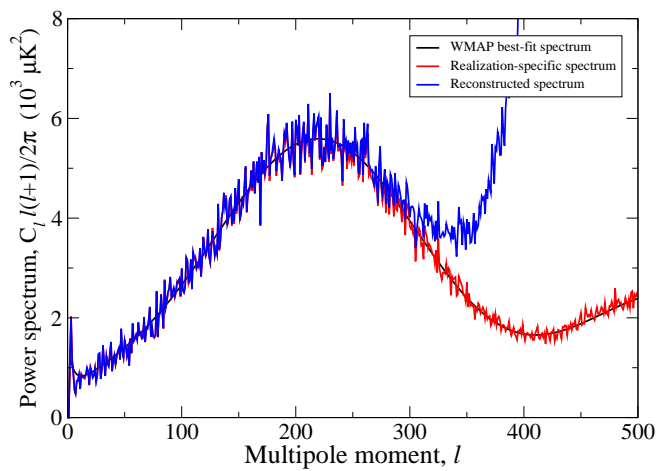}}
\end{center}

\caption{Reconstructed (blue curve) and input (red curve) CMB power
  spectra.  The ensemble-averaged spectrum is shown as a black smooth
  curve.  The reconstructed spectrum was computed by full-sky
  integration without noise weights or sky cut.}
\label{fig:powspec}
\end{figure}

In the right panel of Figure~\ref{fig:error_accuracy} we plot the relative CMB
reconstruction error for the high-resolution map, as we did for the
low-resolution map in the previous section.  Two facts are clear from this plot.
First, the bias is small, indicating that the analytic point estimator is quite
accurate.  Second, the histogram does not match the Gaussian distribution well,
but is rather focused around smaller values.  In other words, the errors are
over-estimated by some small factor by the analytic error propagation formula.
This should not be surprising, given the assumptions that went into those
calculations.  Nevertheless, the estimated errors are in fact of the correct order
of magnitude, and they can therefore be used as a mental guide, although not for
quantitative work.

In Figure~\ref{fig:powspec} we plot the power spectrum of the reconstructed
high-resolution CMB map, the true realization specific input spectrum, and the
ensemble-averaged spectrum.  The reconstructed spectrum was computed by full-sky
integrals without noise weights.  The results are therefore excellent---the
reconstruction is virtually perfect up to $\ell=200$, after which a small noise
term starts to make an impact, before the $1^\circ$ beam renders the
reconstruction arbitrary at $\ell=300$.  From this plot it seems clear that we
were too conservative when choosing a $6^\circ$ beam for the low-resolution
analysis, and that the combined \emph{Planck} and six-year \emph{WMAP} data can
easily handle higher resolutions.

\section{Outstanding problems}

As demonstrated in the previous section, the component separation
method presented in this paper works well on simulated data.  However,
there are a few outstanding issues that we have not taken into
account, but that will have to be addressed prior to analysis of real
data.  The most important of these are gain and zero-point
calibration, dipole corrections, noise correlations, and beam
asymmetries.

First, our algorithm requires all sky maps to be properly calibrated
with respect to gain and zero-point.  Usually, the gain is calibrated
using the CMB dipole, but this can be difficult for channels that are
highly foreground-contaminated.  Also, zero-point calibration is never
easy.

Closely related to these issues is dipole subtraction.  The CMB dipole itself is
hard to observe because of the large Doppler-dipole induced by the motion of the
Solar system through space; it is usually subtracted in the map making process. 
Nevertheless, residual dipoles may cause serious problems for our algorithm
unless accounted for.

Finding reliable calibration methods for each of the above problems is clearly
essential.  Fortunately, the number of degrees of freedom represented by these
issues is quite small, and it may be possible to include them in the analysis by
replacing the signal model $S_{\nu}$ in equation
\ref{eq:chisq} with
\begin{equation}
S_{\nu}(p) \leftarrow
g_{\nu}S_{\nu}(p)+\sum_{\ell=0}^{1}\sum_{m=-\ell}^{\ell}
a_{\ell m,\nu}Y_{\ell m}(p).
\end{equation}
Following a rough calibration with external techniques, one could then use
methods similar to those described in this paper to optimize the gains $g_{\nu}$
and monopole and dipole coefficients $a_{\ell m,\nu}$.

From a conceptual point of view, correlated noise poses a more serious problem.
For \emph{Planck}, for example, the main effect will be to introduce stripes in
the sky maps along the scanning path of the detectors, and locally, this has the
same effect as an overall offset.  Properly speaking, correlated noise is a
problem for map making more than it is for component separation; however, residual
effects can be expected.  Only when actual data are in hand will it be clear how
serious a problem it is.

Finally, in this paper we have assumed that all detectors have identical beam
response functions.  This obviously is not true for any real system, and
corresponding errors are unavoidable.  Fortunately, this is likely to have a
negligible effect on the low-resolution analysis, since we smooth with an
additional degree-scale beam, strongly suppressing small-scale asymmetries.  Only
in the high-resolution analysis is this effect likely to be important.

\section{Discussion}

In this paper, we approach the problem of component separation with
CMB data from the perspective of parameter estimation.  Our goal is to
propagate foreground uncertainties all the way from observed data
through to the final products, most importantly to the CMB power
spectrum and cosmological parameters. This is more easily facilitated
with standard parameter techniques than with image processing
techniques.

We proposed and implemented one particular algorithm for performing this task,
based on multi-frequency parametric model fits established by means of a hybrid
of Markov Chain Monte Carlo and analytic methods.  The method was then shown to
work very well on simulated data, with properties corresponding to those of the
future \emph{Planck} and six-year \emph{WMAP} experiments. 

We also outlined how to propagate the foreground-induced errors to the
CMB power spectrum and cosmological parameter errors, using the output
from the MCMC analysis presented here as the input in a Gibbs sampling
algorithm.  As always, only an actual implementation will prove
whether this method works or not, but the theoretical groundwork
appears to be sound, and no unsurmountable computational problems have
been identified.  Therefore, if this approach proves successful, we
will have a complete, mathematically consistent, end-to-end solution
to the foreground problem in CMB analysis.

While we only considered temperature anisotropy observations in the present
paper, the method is completely general, and can equally well handle polarization
measurements, as will be demonstrated in a future study.  We will also apply the
method to two specific problems.  First, we will study the optimization of
frequency coverage and signal-to-noise ratio in future polarization experiments. 
Since our method provides error bars on all estimated quantities, it is
straightforward to compare different experiment designs.  Modeling errors
will be an integral part of this work, since such uncertainties have a direct
impact on the optimal frequency range to observe.  Second, we will apply
the method to the currently available \emph{WMAP} data.

\begin{acknowledgements}
We thank Jeff Jewell for interesting and useful discussions. HKE
acknowledges financial support from the Research Council of Norway,
including a Ph.\ D.\ studentship. CD thanks Stanley and Barbara Rawn
for funding a fellowship at the California Institute of Technology. We
acknowledge use of the
HEALPix\footnote{http://www.eso.org/science/healpix/} software
\citep{gorski:2005} and analysis package for deriving the results in
this paper.  We also acknowledge use of the Legacy Archive for
Microwave Background Data Analysis (LAMBDA). This work was partially
performed at the Jet Propulsion Laboratory, California Institute of
Technology, under a contract with the National Aeronautics and Space
Administration. CB was supported in part by the NASA Long Term Space
Astrophysics (LTSA) grant NNG04GC90G. EP is an ADVANCE fellow (NSF
grant AST-0340648), also supported by NASA grant NAG5-11489. KMS was
supported by NASA NAG5-10840, the DOE and the Packard Foundation. This
research was supported by a Marie Curie European Reintegration Grant
within the 6th European Community Framework Programme.
\end{acknowledgements}

\end{document}